\def\CarlosI{Instituto de F\'\i sica Te\'orica y Computacional Carlos I, 
Facultad de Ciencias, Universidad de Granada, Campus de Fuentenueva, Granada 18002, 
Spain} 
\def\IAA{Instituto de Astrofisica de Andalucia, Apartado Postal 3004, Granada 
       18080, Spain} 
\def\Comision{Work partially supported by the DGICYT.} 

\def\noi{\noindent}
\def\be{\begin{equation}}
\def\ee{\end{equation}}
\def\bea{\begin{eqnarray}}
\def\eea{\end{eqnarray}}
\def\nn{\nonumber}

\documentclass[11pt,a4paper,twoside]{article}

\usepackage{ihep}

\begin{document}

\begin{center} 
{\large {\bf \uppercase{Lagrangian Formalism on Jet-Gauge\\[1mm]
 and Jet-Diffeomorphism Groups: \\ [1mm]
 Towards a Unification of Gravity \\ [1mm]
 with Internal Gauge Interactions~$^1$} }}
\end{center}

 
\centerline{
{\bf V. Aldaya}~$^{2, \ 3}$ \  and \ {\bf \underline{E. S\'anchez-Sastre}~$^{2,\  3}$} }

\bigskip 
 
\footnotetext[1]{\Comision} 
\footnotetext[2]{\IAA} 
\footnotetext[3]{\CarlosI}

\bigskip 
 
\begin{center} 
\end{center} 
 
\small 
\setlength{\baselineskip}{12pt} 
 
\begin{list}{}{\setlength{\leftmargin}{3pc}\setlength{\rightmargin}{3pc}} 
\item In this talk the description of gauge theories associated with internal symmetries is extended to the case in which the symmetry group is the space-time translation group (recovering Einstein's theory) using the standard jet-bundle formalism. We also reformulate these theories introducing the idea of jet-gauge and jet-diffeomorphism groups.
Finally, we attempt to a simple, 
yet non-trivial, mixing of gravity and electromagnetism (or more general internal interaction) 
by turning to gauge symmetry a central extension of the Poincar\'e group. 

\end{list}



\section{Introduction}

The formulation of gravity as a gauge theory appeared for the first time in Utiyama's works, following an structural scheme analogous to that of internal symmetries which had been introduced by Yang-Mills, and even earlier by Weyl in his attempts to unify electromagnetism and gravitation. After Utiyama's work, there have been many authors (\cite{Kibble, Frolov1, Frolov2, Frolov3, Cho1, Cho2, IVSA, SAZA, Hehl, Hehl2, Hehl3, Hayashi, Basombrio, Andr1, Andr2, Ag1, Ag2}, etc) who have tried to describe gravitation in the framework of gauge interactions. Nevertheless, there is some ambiguity about which external symmetry group is the correct one, i.e. Poincar\'e or just one of its subgroups. Another difference with respect to the internal case is that when dealing with space-time groups it is necessary to introduce new compensating fields $k^{\mu}_{\nu}$, usually known as tetrads or vierbeins, that allow the definition of a non-trivial metric $g_{\mu\nu}$ starting from the flat metric of Minkowski space-time $\eta_{\mu\nu}$.

\noi The traditional formulation of gauge theories is developped by means of the Lagrangian formalism on the 1-jet bundle $J^{1}(E)$ of a bundle E on the Minkowski space-time. It is also assumed as a starting point, the hipothesis of invariance of the action associated with some matter fields under the rigid (global) symmetry Lie group of the free theory. But this procedure would not have a full sense if we desired to study pure radiation from the very begining, since we would not have any matter Lagrangian. Another problem of the usual treatment of gauge theories is related to the interpretation of the gauge potentials as connections, which as is well-known do not transform as a tensor.

\noi In this paper we present gravitation as a consequence of gauging the space-time translation group $T(4)$ with very little deviation from the standard theory. It is worth noting that it is absolutly fair to consider non-trivial translational potentials $\mathcal{A}^{\mu}_{\nu}$ associated with the trivial action of the group $T(4)$ on the fibre, though they are redundant and the constraint $k^{\mu}_{\nu}=\delta^{\mu}_{\nu}+\mathcal{A}^{\mu}_{\nu}$ are required to maintain inaltered the total number of degrees of freedom of the theory. But the relevance of recovering Einstein's equations in the framework of the gauge theory of $T(4)$ is fundamental from the viewpoint of generalizing the conceptual and structural formulation of gauge gravitation theories. In this paper we also present a deep revision of the Minimal Coupling Principle, where the interpretation of gauge fields as connections and the requirement of the matter Lagrangians are removed. The main idea of this reformulation is the concept of ``jet-diffeomorphism group'' of a manifold M, $J^{1}(\mathcal{D}iff(M))$, with coordinates $(\xi^{\mu},\xi^{\mu}_{\nu})$. In this way, the compensating fields (tetrads or vierbeins) $k^{\mu}_{\nu}$ can be interpreted as $k^{\mu}_{\nu}\equiv \xi^{\mu}_{\nu}$, that is, the compensating fields $k^{\mu}_{\nu}$ can be considered as the 1-jets of the diffeomorphims $\xi^{\mu}$ of Minkowski space-time. As a consequence, both indices of $k^{\mu}_{\nu}$ are space-time indices.  
We would like also to remark that we are keeping the name ``gauge'' for the case of the translation group because the general structure of the theory is similar to the structure of gauge theories for internal symmetries. But as was mentioned before, the additional compensating fields $k^{\mu}_{\nu}$ are related to the coordinates of the bundle $J^{1}(\mathcal{D}iff(M))$ and we should properly speak of diffeomorphisms of M.

\noi The version  for internal symmetries arises with the idea of ``jet-gauge group'', 
$G^{1}(M)\equiv J^{1}(G(M))$ with coordinates $(\varphi^{a},\varphi^{b}_{\mu}\equiv A^{(b)}_{\mu})$, where we denote by $G(M)$ the gauge (current or local) group. The gauge potentials $A^{(b)}_{\mu}$ are interpreted as the 1-jets of the Lie group parameters $\varphi^{b}$, departing from the traditional interpretation as connections. Moreover, in this framework the parameters of the Lie group (that is, the functions used to gauge the rigid Lie algebra in the standard formalism) acquire dynamical content.



\noi The paper is organized as follows: Section 2 is devoted to the formulation of the variational principles on the jet bundle 
$J^{1}(E)$. In section 3 we revise the formulation of the gauge theory for internal symmetries and show its corresponding  decription in terms of jet-gauge groups. Section 4 is devoted to the gauge theory of 
the space-time translation group $T(4)$ and the Lagrangian formalism on 
$J^{1}(\mathcal{D}iff(M))$ in order to describe gravity as a variational problem via the Modified Hamilton Principle on the bundle of the 1-jets of the 1-jets of the group of diffeomorphisms of Minkowski space-time 
$J^{1}(J^{1}(\mathcal{D}iff(M)))$.  In section 5, we show an approach to the mixing between electromagnetism and gravity. Finally, we conclude in section 6 and incorporate an appendix about the Lagrangian formalism on  $J^{1}(J^{1}(E))$.



\section{Lagrangian Formalism on the Jet Bundle $J^{1}(E)$}


Classical Field Theory is traditionally formulated on the {\it 1-jet} bundle $J^1(E)$ 
of a vector bundle $E\stackrel{\pi}\rightarrow M$ on Minkowski space-time or any other 
space-time manifold (see e.g. \cite{Hermann,RNC}). The sections of $E$, i.e. mappings $\varphi$ 
from the base manifold $M$ to the total space $E$ such that $\pi\circ\varphi=I_M$, constitute 
the fields, and the 1-jet bundle generalizes to field theory the definition space for 
Lagrangians in ordinary Analytical Mechanics, that is, $T(R^3)\times R\rightarrow R$, 
where $R$ is parametrized by the time $t$ and $T(R^3)$ by $(q^i,\,\dot{q}^j)$. We give 
the formal construction of the jet bundle of a general bundle $E$, non-necessarily a 
vector bundle.

The four first greek letters  
$(\alpha,\beta,\gamma,\delta)$ will denote spinorial indices of the matter fields and the rest of the greek alphabet
$(\epsilon,...,\mu,\nu,...,\tau,...\omega = 0,1,2,3)$ will be used to denote space-time indices.  
Let us consider a  bundle $E$, $\pi: E \rightarrow M$, and name
$\Gamma(E)$ the space of sections of $E$. We define the bundle of the {\it 1-jets} of 
$\Gamma(E)$ as:
\begin{equation}
J^1(E)\equiv\frac{\Gamma(E)\times M}{\stackrel{1}{\sim}}
\end{equation}

\noi where the equivalence relation $\stackrel{1}{\sim}$ is defined as follows:
\begin{equation}
(\varphi,x)\sim(\varphi',x')\Longleftrightarrow\left\{ \begin{array}{rcl}
x&=&x' \\
\varphi(x)&=&\varphi'(x) \\
\partial_{\mu}\varphi(x)&=&\partial_{\mu}\varphi'(x), 
\end{array}\right.\label{relacion}
\end{equation}

\noi $\forall\, (\varphi,x),\,(\varphi',x')\in \Gamma(E)\times M$. The cartesian projection 
$(\varphi,x)\mapsto x$ defines a natural projection $\pi^1:J^1(E)\rightarrow M$.

Locally, we can parametrize $E$ by co-ordinates $(x^\mu,\,\varphi^\alpha)$ defined on $\pi^{-1}(U)$,
$U\subset M$. In fact, let us define the following functions on $\Gamma(\pi^{-1}(U))\times U$:
\bea
\varphi^\alpha:(\varphi,x)&\mapsto&\varphi^\alpha(\varphi(x))\nn\\
\varphi^\alpha_\mu:(\varphi,x)&\mapsto&\partial_\mu\varphi^\alpha(\varphi(x)).
\eea

\noi They are compatible with the equivalence relation $\stackrel{1}{\sim}$ and, therefore, 
go to the quotient defining the corresponding functions $(\varphi^\alpha,\,\varphi^\alpha_\mu)$.

The starting point of the geometric approach to the variational principles is the definition of 
the Lagrangian density as a real function $\mathcal{L}:J^{1}(E)\rightarrow R$ on 
the bundle $J^{1}(E)$ of a vector bundle $E$.  Lagrangians depend locally on the arguments 
$(x^{\mu}, \varphi^{\alpha}, \varphi^{\alpha}_{\mu})$, although usually, due to the Poincar\'e 
invariance, $\mathcal{L}$ will not depend explicitly on $x^{\mu}$.

Let $\varphi$ be a section of $E$. The {\it 1-jet} extension (or prolongation) of $\varphi$, 
$j^{1}(\varphi)\equiv \overline{\varphi}$, is the only section of $J^{1}(E)$ such that $j^{1}$ 
is an injection of $\Gamma(E)$ into $\Gamma(J^{1}(E))$ and $\theta^{\alpha} \mid _{j^{1}(\varphi)(M)}=0$ 
where $\theta^{\alpha}$ are the {\it structure 1-forms} defined on $J^{1}(E)$ by 
\be
\theta^{\alpha}=d\varphi^{\alpha}-\varphi^{\alpha}_{\mu}dx^{\mu}.\label{estructura}
\ee

Given an arbitrary vector field $X$ on $E$, $X\in\Gamma(T(E))\equiv{\cal X}(E)$, 

\begin{equation}
X=X^{\mu}{\large\frac{\!\partial }{\partial x^{\mu}}}+X^{\alpha}{\large\frac{\!\partial }{\partial \varphi^{\alpha}}}\,,
\end{equation}
\noi acting on the space-time and on the fibre, its {\it 1-jet} extension (or prolongation) 
through the injection $j^{1}$ is the 
only vector field $j^{1}(X)\equiv \bar{X}$ on $J^{1}(E)$ such that  $\bar{X}$ projects on X, that is,
\begin{equation}
\bar{X}=X+\bar{X}^{\alpha}_{\mu}{\large\frac{\!\partial }{\partial \varphi^{\alpha}_{\mu}}}\,,
\end{equation}

\noi and it is an infinitesimal contact transformation, i.e.
\begin{equation}
L_{\bar{X}}\theta^{\alpha}=C^{\alpha}_{\beta}\theta^{\beta}\,,
\end{equation}

\noi where $L_{\bar{X}}$ is the Lie derivative with respect to the vector field $\bar{X}$,  and $C^\alpha_\beta$ are constants. 

\noi This second condition implies that 
\bea
C^{\alpha}_{\beta}=\frac{\partial X^{\alpha}}{\partial \varphi^{\beta}}-\varphi^{\alpha}_{\mu}\frac{\partial X^{\mu}}{\partial \varphi^{\beta}},
\eea

\begin{equation}
\bar{X}^{\alpha}_{\mu}= {\large\frac{\partial X^{\alpha} }{\partial x^{\mu}}}-
\varphi^{\alpha}_{\nu}{\large\frac{\partial X^{\nu} }{\partial x^{\mu}}}+\left({\large\frac{\partial X^{\alpha} }
{\partial \varphi^{\beta}}}-\varphi^{\alpha}_{\nu}{\large\frac{\partial X^{\nu} }{\partial \varphi^{\beta}}}\right)
\varphi^{\beta}_{\mu}\,.\label{extjet}
\end{equation}
 

%


Given a Lagrangian density $\mathcal{L}$, the Hamilton functional (the action )\, ${\cal S}$ \, is the 
mapping from $\Gamma(E)$ to $R$ defined by
\[\mathcal{S}:\Gamma(E) \rightarrow R\]

\begin{equation} 
\mathcal{S} (\varphi)=\int_{j^{1}(\varphi)(M)} \mathcal{L}(j^{1}(\varphi)){\pi^1}^*\omega\,,
\end{equation} 

\noi $\forall \varphi \in \Gamma(E)$, where $\omega=dx^{0}\wedge dx^{1}\wedge dx^{2}\wedge dx^{3}$ 
is the volume 4-form on $M$ and ${\pi^1}^*\omega$ its pull-back to $J^1(E)$ (to be denoted just $\omega$ in the sequel).

The Ordinary Hamilton Principle states that the critical sections (trajectories) of the variational 
problem are the solutions of
\[(\delta \mathcal{S})_{j^{1}(\varphi)}(X)\equiv\int_{j^{1}(\varphi)(M)}L_{\bar{X}}(\mathcal{L}(j^{1}
(\varphi))\omega)=0\,,\]

\noi where $X$ is an arbitrary vector field on $E$.

As is well known, this principle leads to the usual Euler-Lagrange motion equations
\begin{equation}
\frac{\partial \mathcal{L}}{\partial \varphi^{\alpha}}-\frac{d }{dx^{\mu}}\left(\frac{\partial \mathcal{L}}
{\partial \varphi^{\alpha}_{\mu}}\right)=0\,. \label{EcLagrange} 
\end{equation}  




The variational calculus can by generalized through the Modified Hamilton Principle, which varies
sections of $J^1(E)$ rather than sections of $E$. To this end a {\it modified Hamiltonian action} 
$\mathcal{S}^1:\Gamma(J^1(E)) \rightarrow R$ is defined as follows:
\be
\mathcal{S}^1(\varphi^{1})=\int_{\varphi^{1}(M)}\Theta_{PC}\,,\label{action'}
\ee 

\noi  where $\Theta_{PC}$, the Poincar\'e-Cartan(-Hilbert) form is a $n\equiv\hbox{dim}M$ form which 
generalizes that of Mechanics, $\frac{\partial L}{\partial\dot{q}^i}(dq^i-\dot{q}^idt)+Ldt=p_idq^i-Hdt$:
\bea
\Theta_{PC}&=&\frac{\partial{\cal L}}{\partial\varphi^\alpha_\mu}(d\varphi^\alpha-\varphi^\alpha_\nu
    dx^\nu)\wedge\theta_\mu+{\cal L}\omega\nn\\
&=&\frac{\partial{\cal L}}{\partial\varphi^\alpha_\mu}d\varphi^\alpha\wedge\theta_\mu-
   \left(\frac{\partial{\cal L}}{\partial\varphi^\alpha_\mu}\varphi^{\alpha}_\mu-{\cal L}\right)\omega\nn\\
&\equiv&\pi^\mu_\alpha d\varphi^\alpha\wedge\theta_\mu-{\cal H}\omega\;, \label{TetaPC}
\eea

\noi where $\theta_\mu\equiv i_{\frac{\!\partial}{\partial x^\mu}}\omega$. In the regularity case, i.e.
$det(\frac{\partial^2{\cal L}}{\partial\varphi_\mu\partial\varphi_\nu})\neq 0$, ${\cal H}$ is the covariant 
Hamiltonian and $\pi^\mu_\alpha$ the $n$ covariant momenta.

The {\it Modified Hamilton Principle} stablishes that the physical 
trajectories are those sections of $J^1(E)$, points in $\Gamma(J^1(E))$, 
where the derivative of the action (\ref{action'}) is zero:
\bea
(\delta\mathcal{S'})_{\varphi^1}(X^1)&\equiv&\int_{\varphi^1(M)}L_{X^1}\Theta_{PC}=0\;,
\;\;\;\forall X^1\in{\cal X}(J^1(E))\nn\\
&\Rightarrow&\;i_{X^1}d\Theta_{PC}|_{\varphi^1}=0.\label{EcLagrange'}
\eea

Equations (\ref{EcLagrange'}) generalize the Euler-Lagrange ones 
(\ref{EcLagrange}) and, in the case of regularity, reproduce them and 
provide the covariant Hamilton equations\cite{RNC}:
\be
\frac{\partial{\cal H}}{\partial\pi^\mu_{\alpha}}=\frac{\partial\varphi^{\alpha}}{\partial x^\mu}\;,
   \;\;\;\frac{\partial{\cal H}}{\partial\varphi^{\alpha}}=-\frac{\partial\pi^\mu_{\alpha}}
    {\partial x^\mu}\;.\label{EcHamilton}
\ee

\bigskip

\section{Gauge Theory of Internal Symmetries and Jet-Gauge Groups}

Let us consider a matter Lagrangian density 
\begin{equation}
\mathcal{L}_{matt}(\varphi^{\alpha},\varphi_{\mu}^{\alpha})
\end{equation}

\noi and the corresponding action 
\begin{equation}
\mathcal{S}=\int\mathcal{L}_{matt}\;\omega\,.
\end{equation}

Let us assume that $\mathcal{S}$ is invariant under a global (or rigid) Lie group of internal 
symmetry G, i.e., 
\begin{equation}
L_{\bar{X}_{(a)}}\left(\mathcal{L}_{matt}\;\omega\right)=0
\end{equation}

\noi where $\bar{X}_{(a)}\equiv j^{1}(X_{(a)})$ is the jet extension of the generator 
\begin{equation}
X_{(a)}=X^{\alpha}_{(a)\beta}\varphi^{\beta}\frac{\!\partial }{\partial \varphi^{\alpha}}
\end{equation}

\noi of G (whose components on $\frac{\!\partial}{\partial x^\mu}$ have been suppresed 
since we have supposed that $G$ does not act on the base manifold $M$) satisfying the commutation relations

\[(X_{(b)}X_{(a)}-X_{(a)}X_{(b)})^{\alpha}_{\beta}=C^{c}_{ab}X^{\alpha}_{(c)\beta}.\]

\noi We use the labels 
$(a),\,(b),...$ to denote group indices. Needless to say that the term 
$\mathcal{L}L_{\bar{X}}\omega=0$ since $div(X)=0$.

According to the general expression for the jet extension and the assumption of the internal action of $G$, 
the invariance of the action under the rigid group G may be written simply as:
\begin{equation}
\bar{X}_{(a)}\mathcal{L}_{matt}(\varphi^{\alpha},\varphi^{\alpha}_{\mu})=
X^{\alpha}_{(a)\beta}\varphi^{\beta}\frac{\partial \mathcal{L}_{matt}}
{\partial \varphi^{\alpha}}(\varphi^{\alpha},\varphi^{\alpha}_{\mu})+
X^{\alpha}_{(a)\beta}\varphi^{\beta}_{\mu}
\frac{\partial \mathcal{L}_{matt}}
{\partial 
\varphi^{\alpha}_{\mu}}(\varphi^{\alpha},\varphi^{\alpha}_{\mu})=0\,.
\end{equation}

Now we wish to study the invariance of the action under the (``current'', ``local'' or) gauge  
group $G(M)$. The corresponding Lie algebra  is the following tensor product: 
\begin{equation}
\mathcal{F}(M)\otimes\mathcal{G}\equiv 
\{f^{(a)}X_{(a)}\}\,,\label{algebragauge}
\end{equation} 

\noi where $\mathcal{F}(M)$ is the multiplicative algebra of real analytic functions on M, and 
$\mathcal{G}$ is the Lie algebra of the Lie group G. Now the jet extension of the generators of 
this new algebra takes the form:
 \begin{equation}
\overline{f^{(a)}X_{(a)}}=f^{(a)}X^{\alpha}_{(a)\beta}\varphi^{\beta}\frac{\!\partial }
{\partial \varphi^{\alpha}}+\left(f^{(a)}X^{\alpha}_{(a)\beta}\varphi^{\beta}_{\mu}+
X^{\alpha}_{(a)\beta}\varphi^{\beta}\frac{\partial f^{(a)} }{\partial x^{\mu}}\right)
\frac{\!\partial }{\partial \varphi^{\alpha}_{\mu}}.
\end{equation}

\noi We see that the jet extension of a generator in (\ref{algebragauge}) is no longer a tensor 
product but, rather, it acquires extra terms:
\begin{equation}
\overline{f^{(a)}X_{(a)}}=f^{(a)}\bar{X}_{(a)}+X^{\alpha}_{(a)\beta}\varphi^{\beta}
\frac{\partial f^{(a)} }{\partial x^{\mu}}\frac{\!\partial }{\partial 
\varphi^{\alpha}_{\mu}}\,.
\end{equation}

\noi Therefore, 

\[\overline{f^{(a)}X_{(a)}}\mathcal{L}_{matt}\neq0\]

\noi and, in order to  restore the invariance, we have to introduce new compensating fields (usually known 
as Yang-Mills fields) $A^{(a)}_{\mu}$ with the following transformation law under $G(M)$: 
\begin{equation}
X_{A^{(a)}_{\mu}}\equiv \delta A^{(a)}_{\mu}=f^{(b)}C^{a}_{bc}A^{(c)}_{\mu}+
{\large\frac{\partial f^{(a)}}{\partial x^{\mu}}}\,,
\end{equation}

\noi where $C^{a}_{bc}$ are the structure constants of the Lie algebra $\mathcal{G}$.

Then, we must consider the generators acting also on the new fields 
$A^{(a)}_\mu$:
\begin{equation}
f^{(a)}\mathcal{X}_{(a)}=f^{(a)}X^{\alpha}_{(a)\beta}\varphi^{\beta}\frac{\!\partial }
{\partial \varphi^{\alpha}}+\left(f^{(b)}C^{a}_{bc}A^{(c)}_{\mu}+\frac{\partial f^{(a)}}
{\partial x^{\mu}}\right)\frac{\!\partial }{\partial A^{(a)}_{\mu}}\,.
\end{equation}

Now the situation is as follows: we have introduced new fields $A^{(a)}_{\mu}$ to modify the 
behaviour of the original Lagrangian of matter so that we have to find,
on the one hand the expression for the new Lagrangian $\widehat{\mathcal{L}}_{matt}$ containing the
matter fields and their interaction with the new compensating fields $A^{(a)}_{\mu}$ and, on 
the other, the free Lagrangian $\mathcal{L}_{0}$ corresponding to the new fields, which should depend 
on the new fields variables and their first derivatives, i.e. $A^{(a)}_{\mu}$, $A^{(a)}_{\nu,\sigma}$.
Let us establish the required process under the aspect of a two-parts theorem to be 
referred to as Utiyama's Theorem (as suggested in mathematical literature) without aiming at a too formal 
presentation. Proofs of theorems will be presented elsewhere.

\bigskip
 
\noi {\bf Theorem 1.A (The Minimal Coupling):} {\it The new Lagrangian describing the matter fields 
as well as their interaction with the new compensating fields $A^{(a)}_{\nu}$},
\begin{equation}
\widehat{\mathcal{L}}_{matt}(\varphi^{\alpha},\varphi^{\alpha}_{\mu},A^{(a)}_{\nu})\equiv 
\mathcal{L}_{matt}(\varphi^{\alpha},\varphi^{\alpha}_{\mu}-A^{(a)}_{\mu}X^{\alpha}_{(a)\beta}\varphi^{\beta}),
\end{equation}

\noi {\it is invariant under the current group} $G(M)$ {\it in the sense that} 
\begin{equation}
\overline{f^{(a)}\mathcal{X}_{(a)}}\widehat{\mathcal{L}}_{matt}(\varphi^{\alpha},
\varphi^{\alpha}_{\mu},A^{(a)}_{\nu})=0\,.
\end{equation} 





\bigskip


Thus, as {\bf Corollary 1.A}, we conclude that the matter fields interact only with the new 
compensating fields and not with their derivatives and 
the actual interaction term appears in the Lagrangian through the combination 
\begin{equation}
\phi^{\alpha}_{\mu}=\varphi^{\alpha}_{\mu}-A^{(a)}_{\mu}X^{\alpha}_{(a)\beta}\varphi^{\beta},
\end{equation} 

\noi usually known as covariant derivative of the matter fields (on this respect, see the end of 
this subsection).


The introduction of the new compensating fields naturally leads to considering a new action 
accounting also for the dynamics of these new fields with Lagrange density $\mathcal{L}_0$: 
\begin{equation}
\mathcal{S'}=\int (\widehat{\mathcal{L}}_{matt}\omega+\mathcal{L}_{0}\omega).
\end{equation}

\noi Since we have already seen that $\;\int \widehat{\mathcal{L}}_{matt}\;\omega\;$ is 
invariant under $G(M)$, imposing the invariance of $\;\mathcal{S'}\;$ requires the invariance of
$\;\;\int \mathcal{L}_{0}\;\omega\;\;$ itself.  That is, the free Lagrangian $\mathcal{L}_{0}$, 
containing the new compensating  fields and their first derivatives, must be invariant under the 
current group $G(M)$: 
\begin{equation}
\overline{f^{(a)}\mathcal{X}_{(a)}}\mathcal{L}_{0}(A^{(c)}_{\mu},A^{(b)}_{\nu,\sigma})=0\,.
\end{equation}

\noi In the proof of the following ``theorem'' we solve this system and obtain the structure 
of $\mathcal{L}_{0}$.

\medskip

\noi{\bf Theorem 1.B (Structure of the free Lagrangian $\mathcal{L}_{0}$):} {\it The necessary 
condition for $\mathcal{L}_0$ to be invariant under the current group $G(M)$ is that $\mathcal{L}_0$
depends on the fields $A^{(a)}_\mu$ and their ``derivatives'' $A^{(a)}_{\mu,\nu}$ only through the 
specific combination:}
\begin{equation}
F^{(a)}_{\mu\nu}\equiv A^{(a)}_{\mu,\nu}-A^{(a)}_{\nu,\mu}-\frac{1}{2}C^{a}_{bc}(A^{(b)}_{\mu}A^{(c)}_
{\nu}-A^{(b)}_{\nu}A^{(c)}_{\mu})\,,\label{curvatura}
\end{equation}

\noi {\it i.e.}
\begin{equation}
\overline{f^{(a)}\mathcal{X}_{(a)}}\mathcal{L}_{0}(A^{(c)}_{\mu},A^{(b)}_{\nu,\sigma})=0
 \Longrightarrow \mathcal{L}_{0}=\mathcal{L}_{0}(F^{(a)}_{\mu\nu})\,.
\end{equation}
  
\noi The expression (\ref{curvatura}) is traditionally called ``curvature'' of the ``connection'' 
$A^{(a)}_\mu$ (see again the end of this subsection).

\medskip

It should be remarked that the actual dependence of $\mathcal{L}_0$ on the tensor $F$ is not 
fixed and must be chosen with the help of extra criteria, for example the invariance under the rigid Poincar\'e group. In particular, to account for the 
standard Yang-Mills equations the Lagrangian must be of the form 
\begin{equation}
\mathcal{L}_{0}\sim \sum_{a=1}^{dimG} F^{(a)}_{\mu\nu}F^{(a)}_{\sigma\rho}\eta^{\sigma\mu}
\eta^{\rho\nu}\,.
\end{equation}

\noi For space-time symmetries the ambiguity in the specific dependence of $\mathcal{L}_0$ will 
be patent.
\bigskip

\noi{\bf Euler-Lagrange equations}

\bigskip

Let us consider the Lagrangian density $\mathcal{L}_{tot}\equiv \widehat{\mathcal{L}}_{matt}
+\mathcal{L}_{0}$, 
where $\widehat{\mathcal{L}}_{matt}(\varphi^{\alpha},\varphi^{\alpha}_{\mu},A^{(a)}_{\nu})\equiv 
\mathcal{L}_{matt}(\varphi^{\alpha},\varphi^{\alpha}_{\mu}-A^{(a)}_{\mu}X^{\alpha}_{(a)\beta}
\varphi^{\beta})$. Then, the equations of 
motion are given by:

\be
\frac{\!d}{dx^\mu}\left(\frac{\partial \mathcal{L}_{matt}}{\partial \varphi^{\alpha}_{\mu}}\right)-
\frac{\partial \mathcal{L}_{matt}}{\partial \varphi^{\alpha}}+ X^{\beta}_{(a)\alpha}A^{(a)}_\mu
\frac{\partial \mathcal{L}_{matt}}{\partial \varphi^{\beta}_{\mu}}=0
\ee
\begin{equation}
\frac{\!d}{dx^\nu}\left(\frac{\partial \mathcal{L}_{0}}{\partial F^{(a)}_{\mu\nu}}\right)+
C^{b}_{da}A^{(d)}_{\nu}\frac{\partial \mathcal{L}_{0}}{\partial F^{(b)}_{\nu\mu}}+
\frac{1}{2}X^{\alpha}_{(a)\beta}\varphi^{\beta}\frac{\partial \mathcal{L}_{matt}}
{\partial \varphi^{\alpha}_{\mu}}=0\,.
\end{equation}

\bigskip

\noi {\bf Remark:} To finish the section corresponding to internal symmetries, we would like to 
note that, on sections, 
\begin{equation}
F^{(a)}_{\mu\nu}=\partial_{\nu}A^{(a)}_{\mu}-\partial_{\mu}A^{(a)}_{\nu}-\frac{1}{2}(A^{(b)}_
{\mu}A^{(c)}_{\nu}-A^{(b)}_{\nu}A^{(c)}_{\mu}).
\end{equation}

\noi Introducing the notation (spin connection)
\[\Gamma^{\alpha}_{\mu\beta}\equiv A^{(a)}_{\mu}X^{\alpha}_{(a)\beta}\,,\]

\noi and taking into account the commutation relations
\[(X_{(b)}X_{(a)}-X_{(a)}X_{(b)})^{\alpha}_{\beta}=C^{c}_{ab}X^{\alpha}_{(c)\beta}\,,\]

\noi the tensor $F^{(a)}_{\mu\nu}$ can be turned into a curvature tensor:
\bea
R^{\alpha}_{\mu\nu\beta}\equiv F^{(a)}_{\mu\nu}X^{\alpha}_{(a)\beta}&=&\partial_{\mu}
\Gamma^{\alpha}_{\nu\beta}-\partial_{\nu}\Gamma^{\alpha}_{\mu\beta}-\frac{1}{2}C^{a}_{bc}
(A^{(b)}_{\mu}A^{(c)}_{\nu}(X_{(a)})^{\alpha}_{\beta}-A^{(b)}_{\nu}A^{(c)}_{\mu}(X_{(a)})^{\alpha}_
{\beta})\nn\\
&=&\partial_{\mu}\Gamma^{\alpha}_{\nu\beta}-\partial_{\nu}\Gamma^{\alpha}_{\mu\beta}-
(\Gamma^{\alpha}_{\mu\gamma}\Gamma^{\gamma}_{\nu\beta}-\Gamma^{\alpha}_{\nu\beta}\Gamma^{\gamma}_
{\mu\beta})\,.
\eea

\noi Accordingly, the Minimal Coupling prescription can be easily reformulated by saying that 
the matter Lagrangian incorporating the interaction terms  is obtained from the original one by 
replacing all derivatives of the matter fields with covariant derivatives, and that the 
Lagrangian for the connections must be a function 
of the curvature. \\[-2mm]

Now we would like to present a conceptual and structural revision of the standard formulation of gauge theories, in such a way that the gauge group acquires dynamical character, and associated natural structures, which substitute the well-known spin connections, appear as physical fields (and these are tensorial objects) irrespective of the existence of fermionic matter.

Let $G$ be a Lie group of internal symmetry, $G(M)$ the gauge group (also named current group), whose elements are the mappings from Minskowski space-time $M$ into $G$ (thus $G(M)$ may be viewed as the set of sections of a principal bundle with structural group $G$). Let $\{x^{\mu}\}$ be a coordinate system on the base manifold $M$ and $\{\varphi^{\alpha}\}$ the corresponding coordinate system on the fibre G.
Let's introduce the bundle of the 1-jets of the gauge group $G(M)$: 
\[J^{1}(G(M)).\] The formal definition of $J^{1}(G(M))$ is given through the relation of equivalence $\sim ^{1}$ in the following way:

\[J^{1}(G(M))\equiv G(M)\times M/\sim ^{1}\]
such that
 \[(\Phi,m)\sim ^{1} (\Phi',m')\] for all $(\Phi,m)$,$(\Phi',m')$ belonging to $G(M)\times M,$ if and only if

1)\[ m=m' \ ,\]

2)\[\Phi(m)=\Phi'(m) \ ,\]

3)\[\partial_{\mu}\Phi(m)=\partial_{\mu}\Phi'(m).\]

This definition may easily be extended to the order r.

A coordinate system for $J^{1}(G(M))$ would be 
\[\{x^{\mu},\varphi^{\alpha},\varphi^{\alpha}_{\mu}\},\] where $\varphi^{\alpha}_{\mu}$ is not necessarily the derivative of any section $\varphi^{\alpha}(x)$. In order to introduce ``derivatives'' of the fields in the theory, we introduce the bundle of the 1-jets of the 1-jets of the gauge group 
\[J^{1}(J^{1}(G(M))),\] defined in an analogous way as $J^{1}(G(M))$ (see the Appendix). 

A suitable coordinate system for $J^{1}(J^{1}(G(M)))$ is, according to the appendinx devoted to the Lagrangian formalisn on $J^{1}(J^{1}(E))$,
 
\[\{x^{\mu},\varphi^{\alpha},\varphi^{\alpha}_{\mu};\varphi^{\alpha}_{,\nu},\varphi^{\alpha}_{\mu,\nu}\}.\]

\noi According to the more usual field theory notation we have replaced the coordinates y and z of the appendix by $\varphi^{\alpha}$ and
$\varphi^{\alpha}_{,}$.


Once we have a certain coodinates system we can construct Lagrangian densities defined on $J^{1}(J^{1}(G(M)))$ and now we will explicitly proceed to study the abelian and non-abelian internal cases.

\subsection{Abelian internal gauge interaction}

The associated Lie group is $U(1)$ and the coordinates that we will use $\{\phi,\phi_{\mu};\phi_{,\nu},\phi_{\mu,\nu}\}$. Due to the invariance under Poincar\'e group the Lagrangian will not explicitly depend on $x^{\mu}$. The more general scalar Lagrangian on $J^{1}(J^{1}(U(1)(M)))$ is given by the following combination with coefficients that will be determined later:

\[\mathcal{L}^{1,1}_{Abel}=a\phi_{\mu,\nu}\phi^{\mu,\nu}+b\phi_{\mu,\nu}\phi^{\nu,\mu}+c\phi^{\mu}_{,\mu}\phi^{\nu}_{,\nu}+e\phi\phi^{\mu}_{,\mu}+f\phi_{\mu}\phi^{\mu}+g\phi_{,\mu}\phi^{,\mu}\]
\[+h\phi_{\mu}\phi^{,\mu}
+j\phi\phi \ .\]

Now we write the motion equations derived from the Modified Hamilton Principle:


\[\delta \phi_{\nu,\rho}:\quad
\frac{\partial}{\partial \phi_{\nu,\rho}}\left(\frac{\mathcal{L}^{1,1}}{\partial \phi_{,\mu}}\right)\left(\frac{\partial \phi}{\partial x^{\mu}}-\phi_{,\mu}\right)+\frac{\partial}{\partial \phi_{\nu,\rho}}\left(\frac{\mathcal{L}^{1,1}}{\partial \phi_{\sigma,\mu}}\right)\left(\frac{\partial \phi_{\sigma}}{\partial x^{\mu}}-\phi_{\sigma,\mu}\right)=0 \ ,\]

\[(a\eta^{\nu\sigma}\eta^{\mu\rho}+b\eta^{\mu\sigma}\eta^{\nu\rho}+c\eta^{\nu\mu}\eta^{\sigma\rho})\left(\frac{\partial \phi_{\nu}}{\partial x^{\mu}}-\phi_{\nu,\mu}\right)=0\]

\[\delta \phi_{,\sigma}:\quad
\frac{\partial}{\partial \phi_{,\sigma}}\left(\frac{\mathcal{L}^{1,1}}{\partial \phi_{,\mu}}\right)\left(\frac{\partial \phi}{\partial x^{\mu}}-\phi_{,\mu}\right)+\frac{\partial}{\partial \phi_{,\sigma}}\left(\frac{\mathcal{L}^{1,1}}{\partial \phi_{\nu,\mu}}\right)\left(\frac{\partial \phi_{\nu}}{\partial x^{\mu}}-\phi_{\nu,\mu}\right)=0\]

\[g\eta^{\sigma\mu}\left(\frac{\partial {\phi}}{\partial x^{\mu}}-\phi_{,\mu}\right)=0 \Longrightarrow \phi_{,\mu}=\partial_{\mu}\phi\]

\[\delta \phi_{\nu}:\quad
-\frac{d}{dx^{\mu}}\left(\frac{\mathcal{L}^{1,1}}{\partial \phi_{\nu,\mu}}\right)+\frac{\partial \mathcal{L}^{1,1}}{\partial \phi_{\nu}}+
\frac{\partial}{\partial \phi_{\nu}}\left(\frac{\mathcal{L}^{1,1}}{\partial \phi_{,\mu}}\right)\left(\frac{\partial \phi}{\partial x^{\mu}}-\phi_{,\mu}\right)
\]
\[+\frac{\partial}{\partial \phi_{\nu}}\left(\frac{\mathcal{L}^{1,1}}{\partial \phi_{\sigma,\mu}}\right)\left(\frac{\partial \phi_{\sigma}}{\partial x^{\mu}}-\phi_{\sigma,\mu}\right)=0\]

\[-2a\partial_{\mu}\phi^{\nu,\mu}-2b\partial_{\mu}\phi^{\mu,\nu}-2c\eta^{\mu\nu}\partial_{\mu}\phi^{\sigma}_{,\sigma}-e\eta^{\mu\nu}\partial_{\mu}\phi+2f\phi^{\nu}+h\phi^{,\nu}+h\eta^{\mu\nu}(\partial_{\mu}\phi-\phi_{,\mu})=0\]

\[\delta \phi:\quad
-\frac{d}{dx^{\mu}}\left(\frac{\mathcal{L}^{1,1}}{\partial \phi_{,\mu}}\right)+\frac{\partial \mathcal{L}^{1,1}}{\partial \phi}+
\frac{\partial}{\partial \phi}\left(\frac{\mathcal{L}^{1,1}}{\partial \phi_{,\mu}}\right)\left(\frac{\partial \phi}{\partial x^{\mu}}-\phi_{,\mu}\right)\]
\[+\frac{\partial}{\partial \phi}\left(\frac{\mathcal{L}^{1,1}}{\partial \phi_{\nu,\mu}}\right)\left(\frac{\partial \phi_{\nu}}{\partial x^{\mu}}-\phi_{\nu,\mu}\right)=0\]

\[-2g\partial_{\mu}\phi^{,\mu}-h\partial_{\mu}\phi^{\mu}+e\phi^{\mu}_{,\mu}+e\eta^{\mu\nu}(\partial_{\mu}\phi_{\nu}-\phi_{\nu,\mu})+2j\phi=0
\ .\]

Making the following choice of coefficients:

$2a=1$

$2f=-m^{2}$

$2b=-1$

$h=e$

$c=0$

$g=\frac{1}{2}$

$2j=-M^{2}$

we arrive at the following equations:

\[\delta \phi_{\sigma,\rho}:\quad
\phi_{\mu,\nu}-\phi_{\nu,\mu}=\partial_{\nu}\phi_{\mu}-\partial_{\mu}\phi_{\nu}\]

\[\delta \phi_{,\sigma}:\quad
\phi_{,\mu}=\partial_{\mu}\phi\]

\[\delta \phi_{\nu}:\quad
-\partial_{\mu}\phi^{\nu,\mu}+\partial_{\mu}\phi^{\mu,\nu}-m^{2}\phi^{\nu}=0 \Longrightarrow \partial_{\mu}(\partial^{\nu}\phi^{\mu}-\partial^{\mu}\phi^{\nu})-m^{2}\phi^{\nu}=0 
\]

\[\Longrightarrow
[\eta_{\mu\nu}(\partial_{\sigma}\partial^{\sigma}+m^{2})-\partial_{\mu}\partial_{\nu}]\phi^{\mu}=0
\ ,\]

\noi where we have used the two previous equations.
This equation contains two variants: Maxwell equations ($m=0$, the electromagnetic field being  $A^{\mu}\equiv \phi^{\mu}$) and Proca equations ($m\neq 0$).
If $m^{2}\neq 0$ then, deriving with respect $x^{\nu}$ the last equation, 

\[\eta_{\mu\nu}(\partial_{\sigma}\partial^{\sigma}+m^{2})\partial^{\nu}\phi^{\mu}-\partial_{\sigma}\partial^{\sigma}\partial_{\mu}\phi^{\mu}=0\]
we get the transversality condition
 
\[\partial_{\mu}\phi^{\mu}=0.\]

And finally,

\[\delta \phi:\quad
\partial_{\mu}\partial^{\mu}\phi+M^{2}\phi=0 \ .\] 

\noi This is the Klein-Gordon equation for the scalar field $\phi$. The role of these extra fields is quite analogous to that of the Goldstone bosons which appear in the Spontaneous Symmetry Breaking mechanism. 

With the additional choice $h=e=0$, the Lagrangian $\mathcal{L}^{1,1}_{Abel}$ takes the form 

\[\mathcal{L}^{1,1}_{U(1)}=\frac{1}{2}(\phi_{\mu,\nu}\phi^{\mu,\nu}-\phi_{\mu,\nu}\phi^{\nu,\mu})-\frac{m^{2}}{2}\phi_{\mu}\phi^{\mu}+\frac{1}{2}\phi_{,\mu}\phi^{,\mu}-\frac{M^{2}}{2}\phi\phi.\]

\subsection{Non-abelian internal gauge interaction}

The coordinate system in this case is denoted by $(\phi^{a},\phi^{a}_{\mu};\phi^{a}_{,\nu},\phi^{a}_{\mu,\nu})$. In analogy to the previous case, let us consider the following Lagrangian on $J^{1}(J^{1}(G(M)))$:

\[\mathcal{L}^{1,1}_{Non-Abel}=\mathcal{L}^{1,1}_{Abel}+\alpha C^{a}_{dc}\phi^{d}_{\mu}\phi^{c}_{\nu}\phi^{\mu,\nu}_{a}+\beta C^{a}_{bc}C_{de,a}\phi^{b}_{\mu}\phi^{c}_{\nu}\phi^{d\mu}\phi^{e\nu} \ ,\]

\noi where

\[\mathcal{L}^{1,1}_{Abel}=\frac{1}{2}(\phi^{a}_{\mu,\nu}\phi^{\mu,\nu}_{a}-\phi^{a}_{\mu,\nu}\phi_{a}^{\nu,\mu})-\frac{m^{2}}{2}\phi^{a}_{\mu}\phi^{\mu}_{a}+\frac{1}{2}\phi^{a}_{,\mu}\phi_{a}^{,\mu}-\frac{M^{2}}{2}\phi^{a}\phi_{a}\]

\noi and $C^{a}_{bc}$ are the structure constants.
Using the Modified Hamilton Principle the motion equations read: 

\[\delta \phi^{a}_{\nu,\rho}:\quad
\frac{\partial}{\partial \phi^{a}_{\nu,\rho}}\left(\frac{\mathcal{L}^{1,1}}{\partial \phi^{b}_{,\mu}}\right)\left(\frac{\partial \phi^{b}}{\partial x^{\mu}}-\phi^{b}_{,\mu}\right)+\frac{\partial}{\partial \phi^{a}_{\nu,\rho}}\left(\frac{\mathcal{L}^{1,1}}{\partial \phi^{b}_{\sigma,\mu}}\right)\left(\frac{\partial \phi^{b}_{\sigma}}{\partial x^{\mu}}-\phi^{b}_{\sigma,\mu}\right)=0\]

\[\partial_{\nu}\phi^{a}_{\mu}-\partial_{\mu}\phi^{a}_{\nu}=\phi^{a}_{\mu,\nu}-\phi^{a}_{\nu,\mu}\]

\[\delta \phi^{a}_{,\sigma}:\quad
\frac{\partial}{\partial \phi^{a}_{,\sigma}}\left(\frac{\mathcal{L}^{1,1}}{\partial \phi^{b}_{,\mu}}\right)\left(\frac{\partial \phi^{b}}{\partial x^{\mu}}-\phi^{b}_{,\mu}\right)+\frac{\partial}{\partial \phi^{a}_{,\sigma}}\left(\frac{\mathcal{L}^{1,1}}{\partial \phi^{b}_{\nu,\mu}}\right)\left(\frac{\partial \phi^{b}_{\nu}}{\partial x^{\mu}}-\phi^{b}_{\nu,\mu}\right)=0\]

\[\eta^{\mu\nu}\delta_{ab}(\partial_{\mu}\phi^{b}-\phi^{b}_{,\mu})=0 \Longrightarrow \partial_{\mu}\phi^{b}=\phi_{,\mu}^{b}\]

\[\delta \phi^{a}_{\nu}:\quad
-\frac{d}{dx^{\mu}}\left(\frac{\mathcal{L}^{1,1}}{\partial \phi^{a}_{\nu,\mu}}\right)+\frac{\partial \mathcal{L}^{1,1}}{\partial \phi^{a}_{\nu}}+
\frac{\partial}{\partial \phi^{a}_{\nu}}\left(\frac{\mathcal{L}^{1,1}}{\partial \phi^{b}_{,\mu}}\right)\left(\frac{\partial \phi^{b}}{\partial x^{\mu}}-\phi^{b}_{,\mu}\right)
\]
\[+\frac{\partial}{\partial \phi^{a}_{\nu}}\left(\frac{\mathcal{L}^{1,1}}{\partial \phi^{b}_{\sigma,\mu}}\right)\left(\frac{\partial \phi^{b}_{\sigma}}{\partial x^{\mu}}-\phi^{b}_{\sigma,\mu}\right)=0\]

\[-\partial_{\mu}\phi^{\nu,\mu}_{a}+\partial_{\mu}\phi^{\mu,\nu}_{a}-\alpha C_{dc,a}\partial_{\mu}(\phi^{d\nu}\phi^{c\mu})-m^{2}\phi^{\nu}_{a}+C^{f}_{dc}\alpha (\delta^{d}_{a}\phi^{\nu,\sigma}_{f}\phi^{c}_{\sigma}+\delta^{c}_{a}\phi^{\mu,\nu}_{f}\phi^{d}_{\mu})+
4\beta C^{g}_{ac}C_{de,g}\phi^{c}_{\sigma}\phi^{d\nu}\phi^{e\sigma}\]
\[+C^{f}_{dc}\alpha (\delta_{fb}\eta^{\sigma\nu}\phi^{c\mu}\delta^{d}_{a}+\delta_{fb}\phi^{d\sigma}\eta^{\mu\nu}\delta^{c}_{a})(\partial_{\mu}\phi^{b}_{\sigma}-\phi^{b}_{\sigma,\mu})=0
\ .\]

Using the equation 
$\partial_{\nu}\phi^{a}_{\mu}-\partial_{\mu}\phi^{a}_{\nu}=\phi^{a}_{\mu,\nu}-\phi^{a}_{\nu,\mu}$,
it follows

\[C^{f}_{dc}\alpha (\delta_{fb}\eta^{\sigma\nu}\phi^{c\mu}\delta^{d}_{a}+\delta_{fb}\phi^{d\sigma}\eta^{\mu\nu}\delta^{c}_{a})(\partial_{\mu}\phi^{b}_{\sigma}-\phi^{b}_{\sigma,\mu})=0\]

and, therefore, the third motion equation reads:

\newpage

\[-\partial_{\mu}\partial^{\mu}\phi^{\nu}_{a}+\partial_{\mu}\partial^{\nu}\phi^{\mu}_{a}-m^{2}\phi^{\nu}_{a}-\alpha C_{dc,a}\partial_{\mu}(\phi^{d\nu}\phi^{c\mu})+\alpha C^{b}_{ac}
(\partial^{\mu}\phi^{\nu}_{b}-\partial^{\nu}\phi^{\mu}_{b})
\phi^{c}_{\mu}+4\beta C^{g}_{ac}C_{de,g}\phi^{c}_{\sigma}\phi^{d\nu}\phi^{e\sigma}=0.\]

Making the choice of coefficients: $\alpha =-1$ and $4\beta =1$, we obtain

\[-\partial_{\mu}\partial^{\mu}\phi^{\nu}_{a}+\partial_{\mu}\partial^{\nu}\phi^{\mu}_{a}-m^{2}\phi^{\nu}_{a}+C_{dc,a}\partial_{\mu}(\phi^{d\nu}\phi^{c\mu})-C^{b}_{ac}
(\partial^{\mu}\phi^{\nu}_{b}-\partial^{\nu}\phi^{\mu}_{b})
\phi^{c}_{\mu}+ C^{g}_{ac}C_{de,g}\phi^{c}_{\sigma}\phi^{d\nu}\phi^{e\sigma}=0.\]

Taking $m=0$, this equation is the Yang-Mills motion equation, if we identify $A^{(a)}_{\mu}\equiv \phi^{a}_{\mu}$ as the gauge fields.

Finally,

\[\delta \phi^{a}:\quad
-\frac{d}{dx^{\mu}}\left(\frac{\mathcal{L}^{1,1}}{\partial \phi^{a}_{,\mu}}\right)+\frac{\partial \mathcal{L}^{1,1}}{\partial \phi^{a}}+
\frac{\partial}{\partial \phi^{a}}\left(\frac{\mathcal{L}^{1,1}}{\partial \phi^{b}_{,\mu}}\right)\left(\frac{\partial \phi^{b}}{\partial x^{\mu}}-\phi^{b}_{,\mu}\right)\]
\[+\frac{\partial}{\partial \phi^{a}}\left(\frac{\mathcal{L}^{1,1}}{\partial \phi^{b}_{\nu,\mu}}\right)\left(\frac{\partial \phi^{b}_{\nu}}{\partial x^{\mu}}-\phi^{b}_{\nu,\mu}\right)=0\]

\[\partial_{\mu}\partial^{\mu}\phi^{a}+M^{2}\phi^{a}=0 \ .\]

This is the Klein-Gordon equation for the scalar fields $\phi^{a}$.



\section{Gauge Theory of the Space-Time Translation Group $T(4)$ and Jet-Diffeomorphism Groups}


\noi Let us consider the matter action

\bea
S=\int \mathcal{L}_{matt}(\varphi^{\alpha},\varphi^{\alpha}_{\mu})d^{4}x
\ ,
\eea

\noi where $\varphi^{\alpha}_{\mu}\equiv \frac{\partial \varphi^{\alpha}}{\partial x^{\mu}}$ on jet sections.

\noi The general form of a generator of an arbitrary Lie algebra $\mathcal{G}$ (of a Lie group $G$) is:

\bea
X=X^{\mu}\frac{\partial }{\partial x^{\mu}}+X^{\alpha}\frac{\partial }{\partial \varphi^{\alpha}}\quad ,
\eea
and its jet extension:

\bea
\overline{X}=X+\overline{X}^{\alpha}_{\mu}\frac{\partial }{\partial \varphi^{\alpha}_{\mu}}\quad ,
\eea
where

\bea
\overline{X}^{\alpha}_{\mu}=\frac{\partial X^{\alpha}}{\partial x^{\mu}}+\frac{X^{\alpha}}{\partial \varphi^{\beta}}\varphi^{\beta}_{\mu}-\left(\frac{\partial X^{\nu}}{\partial x^{\mu}}+\frac{\partial X^{\nu}}{\partial \varphi^{\beta}}\varphi^{\beta}_{\mu}\right)\varphi^{\alpha}_{\nu}\quad .
\eea

\noi The invariance condition of the action under the  Lie group $G$ implies the following invariance condition for the Lagrangian density:

\bea
\overline{X}\mathcal{L}(\varphi^{\alpha},\varphi^{\alpha}_{\mu})+\mathcal{L}(\varphi^{\alpha},\varphi^{\alpha}_{\mu})\partial_{\mu}X^{\mu}=0
\ .\eea

\noi Let us take $G=T(4)$, the space-time translation group. The generators of the translation group are given~by


\bea
X_{(a)}=X^{\mu}_{(a)}\frac{\partial }{\partial x^{\mu}} \ ,
\eea
\bea
X^{\mu}_{(\nu)}=\delta^{\mu}_{\nu} \ ,
\eea

\noi where $(a)=(\nu)$ is the group index.
 
The rigid (global) translational invariance condition reads 

\bea
\partial_{\mu}\mathcal{L}=0,
\eea

\noi that is, the Lagrangian density does not depend explicitly on x.







From the algebra point of view, making the parameters of the group depend on x is equivalent to multiply the generators by arbitrary functions of x :

\bea
X^{gauge}=X^{\mu}(x)\frac{\partial }{\partial x^{\mu}}\equiv f^{\mu}(x)\frac{\partial }{\partial x^{\mu}}
\eea

\bea
\overline{X}^{gauge}=f^{\mu}(x)\frac{\partial }{\partial x^{\mu}}-\partial_{\mu}f^{\nu}(x)\varphi^{\alpha}_{\nu}\frac{\partial }{\partial \varphi^{\alpha}_{\mu}} \ .
\eea

 \noi The invariance is broken and in order to restore it we introduce the compensating fields $k^{\nu}_{\mu}$ (and their inverse $q^{\mu}_{\sigma}$: $k^{\nu}_{\mu}q^{\mu}_{\sigma}=\delta^{\nu}_{\sigma}$), whose transformation law under the local $T(4)$ is given by:

\bea
X_{k^{\nu}_{\mu}}\equiv \delta k^{\nu}_{\mu}=k^{\sigma}_{\mu}\frac{\partial f^{\nu}(x)}{\partial x^{\sigma}}\quad .
\eea

\noi We must consider the generators also acting on the new compensating fields:

\bea
\mathcal{X}^{gauge}=f^{\mu}(x)\frac{\partial }{\partial x^{\mu}}+k^{\sigma}_{\mu}\partial_{\sigma}f^{(\nu)}(x)\frac{\partial }{\partial k^{\nu}_{\mu}}
\eea

\bea
{\overline{\mathcal{X}}}^{gauge}=\mathcal{X}^{gauge}+\overline{X}^{\alpha}_{\mu}\frac{\partial }{\partial \varphi^{\alpha}_{\mu}}+\overline{X}_{k^{\mu}_{\epsilon,\sigma}}\frac{\partial }{\partial k^{\mu}_{\epsilon,\sigma}} \ ,
\eea
 
\noi where

\bea
\overline{X}^{\alpha}_{\mu}\equiv \varphi^{\alpha}_{\nu}\partial_{\mu}f^{\nu}(x)
\eea

\bea
\overline{X}_{k^{\mu}_{\epsilon,\sigma}}\equiv \frac{\partial X_{k^{\mu}_{\epsilon}}}{\partial x^{\sigma}}+\frac{\partial X_{k^{\mu}_{\epsilon}}}{\partial k^{\rho}_{\nu}}k^{\rho}_{\nu,\sigma}-\partial_{\sigma}X^{\rho}k^{\mu}_{\epsilon,\rho}=k^{\nu}_{\epsilon}\partial_{\sigma}\partial_{\nu}f^{\mu}(x)+k^{\nu}_{\epsilon,\sigma}\partial_{\nu}f^{\mu}-k^{\mu}_{\epsilon,\rho}\partial_{\sigma}f^{\rho}(x).
\eea

\noi Note that the expression of $\overline{X}_{k^{\mu}_{\nu,\sigma}}$ arises from the general expresion of $\overline{X}^{\alpha}_{\mu}$.
\\ 

\noi {\bf Theorem 2.A (The Minimal Coupling)}: The new Lagrangian describing the matter fields as well as their interaction with the compensating fields $k^{\nu}_{\mu}$

\bea
\widehat{L}_{matt}(\varphi^{\alpha},\varphi^{\alpha}_{\mu},k^{\nu}_{\mu})\equiv \mathcal{L}_{matt}(\varphi^{\alpha},k^{\nu}_{\mu}\varphi^{\alpha}_{\nu})
\eea
is invariant under the gauge group in the sense that

\bea
{\overline{\mathcal{X}}}^{gauge}\widehat{L}_{matt}(\varphi^{\alpha},\varphi^{\alpha}_{\mu},k^{\nu}_{\mu})=0\quad .
\eea

\noi {\bf Theorem 2.B (The matter action)}: The new action describing the matter fields as well as their interaction with the compensating  fields $k^{\nu}_{\mu}$

\bea
\widehat{S}_{matt}=\int \Lambda\widehat{L}_{matt}(\varphi^{\alpha},\varphi^{\alpha}_{\mu},k^{\nu}_{\mu})d^{4}x
\eea

\noi is invariant under the local $T(4)$, where 

\bea
\Lambda\equiv det(q^{\nu}_{\mu})\quad .
\eea

\noi {\bf Theorem 3.A (The Lagrangian $\mathcal{L}_{0}$ for the compensating fields)}: The necessary condition for $\mathcal{L}_{0}$ to be invariant under the local $T(4)$ is that $\mathcal{L}_{0}$ depends on the fields $k^{\nu}_{\mu}$ and their ``derivatives'' $k^{\nu}_{\mu,\sigma}$ only through the specific combination:

\bea
T^{\sigma}_{\mu\nu}=q^{\sigma}_{\rho}(k^{\rho}_{\nu,\theta} k^{\theta}_{\mu}-k^{\rho}_{\mu,\theta}k^{\theta}_{\nu})
\ .
\eea

\noi {\bf Theorem 3.B (The action for the compensating fields):} The new action describing the compensating fields $k^{\nu}_{\mu}$

\bea
\widehat{S}_{0}=\int {\widehat{\mathcal{L}}}_{0}(k^{\nu}_{\mu},k^{\nu}_{\mu,\sigma})\equiv \int \Lambda\mathcal{L}_{0}(T^{\nu}_{\mu\sigma})d^{4}x 
\eea

\noi is invariant under the local $T(4)$\ .
\\

\noi {\bf Conclusion:}

Starting with an invariant action under the rigid $T(4)$

\bea
S=\int \mathcal{L}_{matt}(\varphi^{\alpha},\varphi^{\alpha}_{\mu})d^{4}x
\eea

\noi we have proved that the new action

\bea
\widehat{S}=\int \widehat{\mathcal{L}}(\varphi^{\alpha},\varphi^{\alpha}_{\mu},k^{\mu}_{\nu},k^{\mu}_{\nu,\sigma})d^{4}x
\eea

\noi is invariant under the local $T(4)$, where

\bea
\widehat{\mathcal{L}}(\varphi^{\alpha},\varphi^{\alpha}_{\mu},k^{\mu}_{\nu},k^{\mu}_{\nu,\sigma})={\widehat{\mathcal{L}}}_{matt}(\varphi^{\alpha},\varphi^{\alpha}_{\mu},k^{\mu}_{\nu})+{\widehat{\mathcal{L}}}_{0}(k^{\mu}_{\nu},k^{\mu}_{\nu,\sigma})
\eea

\noi with

\bea
{\widehat{\mathcal{L}}}_{matt}(\varphi^{\alpha},\varphi^{\alpha}_{\mu},k^{\nu}_{\mu})\equiv \Lambda {\mathcal{L}}_{matt}(\varphi^{\alpha},k^{\mu}_{\sigma}\varphi^{\alpha}_{\mu})\
,\eea

\bea
{\widehat{\mathcal{L}}}_{0}(k^{\mu}_{\nu},k^{\mu}_{\nu,\sigma})\equiv \Lambda {\mathcal{L}}_{0}(T^{\mu}_{\nu\sigma}) \ ,
\eea

\bea
\Lambda=det(q^{\nu}_{\mu})\ .
\eea

\noi {\bf Remark:} We can introduce in the theory gauge fields $\mathcal{A}^{(\mu)}_{\nu}$ associated with the action of the translations generators on the fibre. Nevertheless the matter fields components are inaltered under $T(4)$ and we can impose the constraint $k^{\mu}_{\nu}=\delta^{\mu}_{\nu}+\mathcal{A}^{\mu}_{\nu}$, so that the total number of degrees of freedom does not change. 

\noi {\bf Geometrical interpretation:}

\noi Using the compensating fields $k^{\nu}_{\mu}$ we can construct the metric tensor 

\bea
g_{\mu\nu}\equiv q^{\rho}_{\mu}q^{\sigma}_{\nu}\eta_{\rho\sigma}
\eea

\noi with Levi-Civita connection

\bea
\Gamma^{\sigma (L-C)}_{\mu\nu}\equiv \frac{1}{2}g^{\sigma\rho}(\partial_{\mu}g_{\rho\nu}+\partial_{\nu}g_{\rho\mu}-\partial_{\rho}g_{\mu\nu})
\eea

\noi and the curvature tensor of this connection

\bea
R^{\sigma (L-C)}_{\mu\rho\nu}\equiv \partial_{\rho}\Gamma^{\sigma (L-C)}_{\mu\nu}-\partial_{\nu}\Gamma^{\sigma (L-C)}_{\mu\rho}+\Gamma^{\sigma (L-C)}_{\epsilon\rho}\Gamma^{\epsilon (L-C)}_{\mu\nu}-\Gamma^{\sigma (L-C)}_{\epsilon\mu}\Gamma^{\epsilon (L-C)}_{\mu\rho} \ .
\eea

\noi We can decompose the Levi-Civita connection in two parts:

\bea
\Gamma^{\sigma (L-C)}_{\mu\nu}=\Gamma^{\sigma}_{\mu\nu}-K^{\sigma}_{\mu\nu}
\ ,
\eea

\noi where

\bea
\Gamma^{\sigma}_{\mu\nu}\equiv k^{\sigma}_{\rho}q^{\rho}_{\mu,\nu}=-q^{\rho}_{\mu}k^{\sigma}_{\rho,\nu}
\eea

\noi is not a Levi-Civita connection but a Cartan connection. The other term is known as {\it contortion} and is given by

\bea
K^{\sigma}_{\mu\nu}\equiv \frac{1}{2}(\mathcal{T}^{\; \sigma}_{\mu \ \nu}+\mathcal{T}^{\; \sigma}_{\nu \ \mu}-\mathcal{T}^{\sigma}_{\mu\nu})
\eea

\noi with

\bea
\mathcal{T}^{\sigma}_{\mu\nu}=k^{\sigma}_{\rho}q^{\epsilon}_{\mu}q^{\xi}_{\nu}T^{\rho}_{\epsilon\xi}=\Gamma^{\sigma}_{\mu\nu}-\Gamma^{\sigma}_{\nu\mu}.
\eea

\noi $\mathcal{T}^{\sigma}_{\mu\nu}$ can be interpreted as the torsion of the connection $\Gamma^{\sigma}_{\mu\nu}$.

\noi Now we can see that the Hilbert-Einstein Lagrangian density, which depends on the scalar curvature of the Levi-Civita connection, is equivalent to a certain choice of Lagrangian $\mathcal{L}_{0}^{T(4)}$ of the compensating fields $k^{\nu}_{\mu}$ in the gauge theory of the translation group:

\bea
\mathcal{L}_{Hilbert-Einstein}\equiv \Lambda R^{(L-C)}\equiv \Lambda g^{\rho\nu}R^{\mu (L-C)}_{\rho\mu\nu}=\Lambda \mathcal{L}_{0}^{ T(4)}+divergence \ ,
\eea

\noi where

\bea
\mathcal{L}_{0}^{ T(4)}=T^{\mu}_{\nu\sigma}T^{\rho}_{\epsilon\theta}\left(\frac{1}{4}\eta^{\epsilon\nu}\eta^{\sigma\theta}\eta_{\mu\rho}+\frac{1}{2}\eta^{\theta}_{\mu}\eta^{\nu\epsilon}\eta^{\sigma}_{\rho}-\eta^{\sigma}_{\mu}\eta^{\theta}_{\rho}\eta^{\nu\epsilon}\right)
 \ ,\eea

\noi which turns out to be the Lagrangian of the teleparallel description of gravity. Therefore we finally regain Einstein equations:

\bea
R^{(L-C)}_{\mu\nu}-\frac{1}{2}g_{\mu\nu}R^{(L-C)}=0\quad .
\eea


Now we would like to present the corresponding jet-diffeomorphism formalism. We are allowed to define the bundle of the 1-jets of the group of diffeomorphisms 
\bea
J^{1}(\mathcal{D}iff(M))\equiv \frac{\mathcal{D}iff(M)\times M}{\sim^{1}}\quad ,
\eea
\noi where the equivalence relation is defined as follows:

\[1) \quad  m=m' \ ,\]
\[2) \quad  \xi^{\mu}(m)=\xi'^{\mu}(m) \ , \]
\[3) \quad  \partial_{\nu}\xi^{\mu}(m)=\partial_{\nu}\xi'^{\mu}(m) \ ,\] 

\noi $\forall (\xi^{\mu},m),(\xi'^{\mu},m') \in \mathcal{D}iff(M)\times M$.


\noi A coordinate system for $J^{1}(\mathcal{D}iff(M))$ is
$\{\xi^{\mu},\xi^{\mu}_{\nu}\}$.
 Note that, as usual in jet theory, $\xi^{\mu}_{\nu}$ parametrize transformations in the tangent bundle of the base manifold $M$, $T(M)$, which are not the derivative of the diffeomorphism $\xi^{\mu}$ of M, except for jet extensions. The quantization of gravity requires the elimination of the diffeomorphims constraints, so  this parametrization will be very relevant at quantizing because the jets $\xi^{\mu}_{\nu}$ are not simple changes of coordinates, i.e., they are not ``gauge'' in the strict sense of leaving the solution manifold invariant pointwise.
At this point we would like to remark that the structure just defined, $J^{1}(\mathcal{D}iff(M))$, can be identified with the frame bundle with the Minkowski space-time as base manifold. The general definition of the frame bundle with an arbitrary n-dimensional base manifold M is given by the following total bundle space
\bea
F(M)=\{(m,a_{1},a_{2},...,a_{n})/ m \in M,(a_{1},a_{2},...,a_{n}) \,basis\, of\, T_{m}(M)\} \ ,
\eea
\noi where $T_{m}(M)$ is the tangent space of $M$ in the point $m$.

\noi Note that the structure group of this bundle is the linear general group $GL(n)$. This fact is the reason to take into account the gauge theory of this group among the possible gauge theories for gravity, as can be seen in the literature.

\noi It is also worth noting that given a group, the 1-jets of this group also has group structure in general.

\noi In order to allow in the theory the presence of the derivatives of the previous coordinates $\{\xi^{\mu},\xi^{\mu}_{\nu}\}$, we define the bundle of the 1-jets of the 1-jets of the group of diffeomorphisms 
\[J^{1}(J^{1}(\mathcal{D}iff(M))),\] 
\noi with associated coordinate system 
\bea
\{\xi^{\mu},\xi^{\mu}_{\nu};\xi^{\mu}_{,\sigma},\xi^{\mu}_{\nu,\sigma}\}.
\eea

\noi where in general $\xi^{\mu}_{,\sigma}\neq \partial_{\sigma}\xi^{\mu}$ and
$\xi^{\mu}_{\nu,\sigma}\neq \partial_{\sigma}\xi^{\mu}_{\nu}$. 
In this framework the compensating fields $k^{\mu}_{\nu}$, which were introduced in the gauge theory of the space-time translation group in the previous section, may be now viewed as the jets of the diffeomorphisms of Minkowski space-time $\xi^{\mu}_{\nu}$:
\[\xi^{\mu}_{\nu}\equiv k^{\mu}_{\nu}.\]

\noi Then the Minkowski metric $\eta_{\mu\nu}$ can be transformed with $\xi^{\mu}_{\nu}\equiv k^{\mu}_{\nu}$ and a metric $g_{\mu\nu}$ can be defined and thanks to the fact that the gauge theory of the translation group yields Einstein's theory, the same result follows in this construction.

\section{Mixing of Electromagnetism and Gravity}

The simplest procedure to approach the non-trivial mixing between electromagnetism and gravity is 
to consider the central extension of Poincar\'e group as a rigid symmetry, $\mathcal{P}$, by 
the group $U(1)$: 
\begin{equation}
G=\mathcal{P} \widetilde{\otimes} U(1)\equiv \widetilde{\mathcal{P}}\;.
\end{equation}
 
\noi Then the commutator of boosts and translations is modified in such a form that 
\begin{equation}
[K^{i},P_{j}]=\delta^{i}_{j}\left(\frac{1}{c^{2}}P_{0}+\Xi\right)\,,
\end{equation}
 
\noi where $\Xi$ is the generator of $U(1)$.
The group index $(a)$ now runs over $\{(\mu)$ translations, $(\nu\sigma)$ Lorentz,
 $(elec)$ $U(1)\}$. As a report on central extensions of Lie groups with trivial cohomology, i.e. 
pseudo-extensions, and the role that they play in representation theory, we refer the reader to 
Ref. \cite{Pseudo} and references there in.

According to \cite{mezclainfinita} the Lagrangian for the free 
compensating fields should be a general function of the form
\begin{equation}
\mathcal{L}_{0}=\mathcal{L}_{0}(\mathcal{F}^{(\sigma)}_{\mu\nu},\,\mathcal{F}^{(\sigma\rho)}_{\mu\nu},
\,\mathcal{F}^{(elec)}_{\mu\nu})\,,
\end{equation}

\noi where the presence of the mixing constant 
\be
C^{elec}_{\mu,\sigma\rho}\equiv -\kappa (\eta_{\rho\mu}\delta^{0}_{\sigma}-\eta_{\sigma\mu}\delta^{0}_{\rho})\label{Cmixing}
\ee

\noi in the electromagnetic field strength, due to the central pseudo-extension, is to be 
remarked, $\kappa$ being the coupling constant of the mixing.

Let us suppose that the Lagrangian for the free compensating fields has the form:

\bea
\mathcal{L}_{0}=\frac{1}{2}\Lambda (\mathcal{F}^{(ele)}_{\mu\nu}\mathcal{F}^{(ele)\mu\nu}+\mathcal{F}^{(\mu\nu)}_{\mu\nu})\,,
\eea

\noi where $\;\mathcal{F}^{(ele)\mu\nu}\equiv \mathcal{F}^{(ele)}_{\sigma\rho}\eta^{\sigma\mu}\eta^{\rho\nu}\;$, and write the curvatures of the compensating fields in the following way:

\[\mathcal{F}^{(a)}_{\mu\nu}\equiv k^{\sigma}_{\mu}k^{\rho}_{\nu}F^{(a)}_{\sigma\rho}\,,\]

\noi where

\[F^{(a)}_{\sigma\rho}\equiv A^{(a)}_{\sigma,\rho}-A^{(a)}_{\rho,\sigma}+\frac{1}{2}\widetilde{C^{a}_{bc}}(A^{(b)}_{\sigma}A^{(c)}_{\rho}-A^{(b)}_{\rho}A^{(c)}_{\sigma})\,,\]

\[A^{(a)}_{\mu}\equiv q^{\nu}_{\mu}\mathcal{A}^{(a)}_{\nu}\,.\]

\noi Here, $(a)$ runs over the entire group $\widetilde{\mathcal{P}}$ and $\widetilde{C^{a}_{bc}}$
denotes its structure constants including the mixing constant (\ref{Cmixing}).
Explicitly:

\bea
F^{(\epsilon\rho)}_{\mu\nu}&=&A^{(\epsilon\rho)}_{\mu,\nu}-A^{(\epsilon\rho)}_{\nu,\mu}-
\eta_{\theta\sigma}(A^{(\epsilon\theta)}_{\mu}A^{(\sigma\rho)}_{\nu}-
A^{(\epsilon\theta)}_{\nu}A^{(\sigma\rho)}_{\mu})\,,\nn\\
F^{(ele)}_{\mu\nu}&=&A^{(ele)}_{\mu,\nu}-A^{(\epsilon\gamma)}_{\nu,\mu}-
\frac{1}{2}C^{ele}_{\epsilon,\theta\rho}(A^{(\epsilon)}_{\mu}A^{(\theta\rho)}_{\nu}-
A^{(\epsilon)}_{\nu}A^{(\theta\rho)}_{\mu})\,,\nn
\eea

\noi with $A^{(\epsilon\rho)}_{\mu}\equiv q^{\nu}_{\mu}\mathcal{A}^{(\epsilon\rho)}_{\nu}$ and $A^{(\epsilon)}_{\theta}\equiv q^{\nu}_{\theta}\mathcal{A}^{(\epsilon)}_{\nu}=q^{\nu}_{\theta}(k^{\epsilon}_{\nu}-\delta^{\epsilon}_{\nu})=\delta^{\epsilon}_{\theta}-q^{\epsilon}_{\theta}$.

We have not considered $F^{(\sigma)}_{\mu\nu}$, corresponding to the subgroup of translations, because the Lorentz curvature is enough to describe pure gravity in vacuum.

The Euler-Lagrange motion equations read:

\be
(1):\;\;\;\;\; \frac{\partial \mathcal{L}_{0}}{\partial A^{(\nu\rho)}_{\mu}}-\frac{d}{dx^{\sigma}}\left(\frac{\partial \mathcal{L}_{0}}{\partial A^{(\nu\rho)}_{\mu,\sigma}}\right)=0\;\;\Rightarrow \label{(1)} 
\label{(1)}
\ee

\[C^{ele}_{\sigma,\epsilon\theta}A^{(\sigma)}_{\nu}F^{(ele)\mu\nu}+k^{\mu}_{\rho}T^{\rho}_{\epsilon\theta}-k^{\mu}_{\theta}T^{\rho}_{\epsilon\rho}+k^{\mu}_{\epsilon}T^{\rho}_{\theta\rho}+
(k^{\mu}_{\rho}k^{\nu}_{\theta}-k^{\mu}_{\theta}k^{\nu}_{\rho})A^{(\rho}_{\epsilon)\nu}-(k^{\mu}_{\epsilon}k^{\nu}_{\rho}+k^{\mu}_{\rho}k^{\nu}_{\epsilon})A^{(\rho}_{\theta)\nu}=0\,,\]

\noi where

\[F^{(ele)\mu\nu}=F^{(ele)}_{\rho\lambda}g^{\rho\mu}g^{\lambda\nu}\,,\]

\[g^{\rho\mu}=k^{\rho}_{\sigma}k^{\mu}_{\theta}\eta^{\sigma\theta}\,,\]

\[A^{(\mu}_{\nu)\sigma}\equiv \eta_{\nu\rho}A^{(\mu\rho)}_{\sigma}\,,\]

\[T^{\rho}_{\epsilon\theta}=q^{\rho}_{\mu}(k^{\mu}_{\epsilon,\tau}k^{\tau}_{\theta}-k^{\mu}_{\theta,\tau}k^{\tau}_{\epsilon})\;;\]
\be
(2):\;\;\;\;\;  \frac{\partial \mathcal{L}_{0}}{\partial A^{(ele)}_{\mu}}-\frac{d}{dx^{\sigma}}\left(\frac{\partial \mathcal{L}_{0}}{\partial A^{(ele)}_{\mu,\sigma}}\right)=0\;\;\Rightarrow \label{(2)}
\ee

\[\frac{d}{dx^{\sigma}}(\Lambda F^{(ele)\mu\sigma})=0\;;\]

\be
(3):\;\;\;\;\;\;\;  \frac{\partial \mathcal{L}_{0}}{\partial k^{\mu}_{\nu}}-\frac{d}{dx^{\sigma}}\left(\frac{\partial \mathcal{L}_{0}}{\partial k^{\mu}_{\nu,\sigma}}\right)=0\;\;\Rightarrow \label{(3)}
\ee

\[F^{(\nu\sigma)}_{\mu\sigma}-\frac{1}{2}\delta^{\nu}_{\mu}F^{(\sigma\lambda)}_{\sigma\lambda}=\mathcal{T}^{\nu}_{\mu}\,,\]

\noi where

\[\mathcal{T}^{\nu}_{\mu}\equiv \mathcal{T}^{\nu(mix)}_{\mu}+\mathcal{T}^{\nu(ele)}_{\mu}\,.\]

\noi The tensor

\[\mathcal{T}^{\nu(ele)}_{\mu}\equiv -F^{(ele)\nu}_{\sigma}F^{(ele)\sigma}_{\mu}+\frac{1}{2}\delta^{\nu}_{\mu}F^{(ele)}_{\sigma\lambda}F^{(ele)\sigma\lambda}\]

\noi defines the energy-momentum tensor corresponding to the electromagnetic field (with $F^{(ele)\nu}_{\sigma}=g^{\lambda\nu}F^{(ele)}_{\sigma\lambda}$), and

\[\mathcal{T}^{\nu(mix)}_{\mu}=\frac{1}{2}\delta^{\sigma}_{\mu}C^{ele}_{\sigma,\theta\epsilon}q^{\nu}_{\rho}F^{(ele)\rho\tau}A^{(\theta\epsilon)}_{\tau}\;\]

\noi is a direct consequence of the mixing, and it is quite particular, since in the 
absence of matter and considering the mixture between electromagnetism and gravity, we 
would have a non-trivial  energy-momentum tensor.



Equation (\ref{(1)}) gives the form of the Lorentz potentials in terms of $k_{\mu}^{\nu}$, $q_{\rho}^{\sigma}$ and their derivatives. 
Note that for the gauge theory of Poincar\'e group in vacuum the Euler-Lagrange equation for the Lorentz potentials reads:
\[k^{\mu}_{\rho}T^{\rho}_{\epsilon\theta}-k^{\mu}_{\theta}T^{\rho}_{\epsilon\rho}+k^{\mu}_{\epsilon}T^{\rho}_{\theta\rho}+
(k^{\mu}_{\rho}k^{\nu}_{\theta}-k^{\mu}_{\theta}k^{\nu}_{\rho})A^{(\rho}_{\epsilon)\nu}-(k^{\mu}_{\epsilon}k^{\nu}_{\rho}+k^{\mu}_{\rho}k^{\nu}_{\epsilon})A^{(\rho}_{\theta)\nu}=0\,,\]
\noi whose solution is given in equation $(107)$.
In \cite{Kibble} the Lorentz potentials, in the case in which the existence of matter is also considered, are explicitly written (under some assumptions) in terms of the tetrad fields and their derivatives, and using similar steps one could obtain the explicit form of $A^{(\mu\nu)}_{\sigma}$ to the first order in $\kappa$, but for the moment we shall not use this explicit result.

Equation (\ref{(2)}) can be rewritten in the form of Maxwell equations in a gravitational field, but containing  an 
electromagnetic current $J^{\mu(mix)}$ made of gravitational potentials:

\[\partial_{\sigma}[\Lambda g^{\nu\mu}g^{\epsilon\sigma}(A^{(ele)}_{\nu,\epsilon}-A^{(ele)}_{\epsilon,\nu})]=J^{\mu(mix)}\,,\]

\noi with

\[J^{\mu(mix)}\equiv \frac{1}{2}C^{ele}_{\rho,\theta\xi}\partial_{\sigma}[\Lambda g^{\nu\mu}g^{\epsilon\sigma}(A^{(\rho)}_{\nu}A^{(\theta\xi)}_{\epsilon}-A^{(\rho)}_{\epsilon}A^{(\theta\xi)}_{\nu})]
\ .\]

\noi Note that if we do not consider the mixing between electromagnetism and gravity, the 
previous motion equations  reduce to those of the case of gauging Poincar\'e group in vacuum.

In order to study the solution of the equations with mixture a good aproximation will be working 
at first order in the mixing coupling constant $\kappa$ since it is expected that its value is 
about $10^{-8}$ times the electron mass \cite{egmixing}.

The form of these equations is rather complicated, so that we shall restrict our attention 
here to the equation for the electromagnetic field:

\[\partial_{\sigma}(\Lambda F^{(ele)\mu\sigma})=0\;.\]
 
Let us write the electromagnetic field of the theory with mixture to the first order in $\kappa$:

\[A^{(ele)}_{\mu}=A^{(ele)0}_{\mu}+\kappa A^{(ele)1}_{\mu}\,,\]

\noi and, for simplicity, the rest of the compensating fields will be considered  to zero order  in $\kappa$.

Order by order in $\kappa$, the electromagnetic motion equation leads to:\\

\noi {\it Zero order}:
\be
\partial_{\sigma}[\Lambda g^{\mu\epsilon}g^{\sigma\theta}(A^{(ele)0}_{\epsilon,\theta}-A^{(ele)0}_{\theta,\epsilon})]=0\,,
\ee

\noi which shows that $A^{(ele)0}_{\mu}$ is the usual electromagnetic potential in the presence 
of a gravitational field, without mixing.\\

\noi {\it First order}:
\be
\partial_{\sigma}\{\Lambda g^{\mu\epsilon}g^{\sigma\theta}[A^{(ele)1}_{\epsilon,\theta}-A^{(ele)1}_{\theta,\epsilon}-\frac{1}{2}M^{ele}_{\rho,\nu\xi}(A^{(\rho)}_{\epsilon}A^{(\nu\xi)}_{\theta}-A^{\rho}_{\theta}A^{(\nu\xi)}_{\epsilon})]\}=0 \label{(333)}\,, 
\ee

\[M^{ele}_{\rho,\nu\xi}\equiv -\frac{1}{\kappa}C^{ele}_{\rho,\nu\xi}\,.\]

\noi Here the Lorentz potentials $A^{(\mu\nu)}_{\sigma}$ coincide with the Lorentz potentials of 
the case of gauging the Poincar\'e group in vacuum, 
$\mathcal{A}^{Poinc.}_{(\sigma\rho)\mu}=\frac{1}{2}T_{\mu\sigma\rho}+\frac{1}{2}(T_{\sigma\rho\mu}-
T_{\rho\sigma\mu})$ , since we are working to the first order in $\kappa$.

Note that a simple ``mathematical'' solution for $A^{(ele)1}_{\mu,\nu}$ would be
\be
A^{(ele)1}_{\mu,\nu}=\frac{1}{2}M^{ele}_{\rho,\epsilon\theta}A^{(\rho)}_{\mu}A^{(\epsilon\theta)}_{\nu}=\eta_{\sigma\rho}A^{(\rho)}_{\mu}A^{(0\sigma)}_{\nu}=\eta_{ij}A^{(j)}_{\mu}A^{(0i)}_{\nu}
\, ,\label{A1}
\ee

\noi where the last equality follows from the antisymmetry of the Lorentz potentials $A^{(\mu\nu)}_{\sigma}=-A^{(\nu\mu)}_{\sigma}$, and the latin index runs from 1 to 3.
\noi Just having a look at this result, we conclude that the electromagnetic field of the 
mixture theory between electromagnetism and gravity in the context of gauging the central 
(pseudo-)extension of Poincar\'e group by $U(1)$, contains an additional contribution made 
of pure gravitational potentials and this could be the origin of some electromagnetic 
force associated with (very massive) rotating systems, as $A^{(0i)}_\nu$ is somehow 
associated with ``Coriolis-like forces''.

\subsection{Particle in the presence of the mixed electromagnetic and gravitational fields}

Now we would like to consider some situations of physical interest in order to proceed further 
in the understanding and consequences of the mixing between electromagnetism and gravity. 
A natural example is the ``geodesic'' motion.

We shall apply the Minimal Coupling Prescription (developped in the previous sections) to 
the Lagrangian of a particle of mass $m$, momentum
$p_{\mu}(=mu_{\mu}=m\frac{dx_{\mu}}{d\tau})$ and charge $e$:

\bea
\mathcal{L}_{matt}=\frac{1}{2m}p_{\mu}p_{\nu}\eta^{\mu\nu}.
\eea

\noi Now we have to replace $p_{\mu}$ by $k^{\nu}_{\mu}(p_{\mu}-eA^{(ele)}_{\mu})$. Note 
that we are assuming that the particle does not have spin, that is why the term associated 
with the Lorentz subgroup does not contribute to the ``covariant derivative''. The modified
Lagrangian reads:

\[{\widehat{\mathcal{L}}}_{matt}=\frac{1}{2m}k^{\mu}_{\nu}(p_{\mu}-eA^{(ele)}_{\mu})k^{\rho}_{\sigma}(p_{\rho}-eA^{(ele)}_{\rho})\eta^{\nu\sigma}=
\frac{1}{2m}g^{\mu\nu}(p_{\mu}-eA^{(ele)}_{\mu})(p_{\nu}-eA^{(ele)}_{\nu})=\]
\[\frac{m}{2}u^{\mu}u^{\nu}g_{\mu\nu}-eu^{\mu}A^{(ele)\nu}g_{\mu\nu}+\frac{e^{2}}{2m}A^{(ele)\mu}A^{(ele)\nu}g_{\mu\nu}\,,\]

\noi although in what follows we shall neglect the misleading term $\frac{e^{2}}{2m}A^{(ele)\mu}A^{(ele)\nu}g_{\mu\nu}$, which, by the way, does not appear when working directly with the Poincar\'e-Cartan form instead of with the Lagrangian.

We also consider the Lagrangian for the free compensating fields $\mathcal{L}_{0}(\mathcal{F}^{(\sigma)}_{\mu\nu},\mathcal{F}^{(\epsilon\theta)}_{\mu\nu},\mathcal{F}^{(ele)}_{\mu\nu})$ and the action \, $S=\int \mathcal{L}_{Tot}d^{4}x$ \, with \, $\mathcal{L}_{Tot}\equiv {\widehat{L}}_{matt}+\Lambda \mathcal{L}_{0}(\mathcal{F}^{(\sigma)}_{\mu\nu},\mathcal{F}^{(\epsilon\theta)}_{\mu\nu},\mathcal{F}^{(ele)}_{\mu\nu})$,\,  invariant under the local group $\widetilde{\mathcal{P}}(M)$.

As regards the interaction between a particle and a field, in general, it is required to 
distinguish between the coordinates $y^{\sigma}$ where the fields are evaluated and the 
coordinates $x^{\sigma}$ for the particle.
In ${\widehat{\mathcal{L}}}_{matt}$ the fields are evaluated at the position of the particle, 
where the interaction occurs, but in $\mathcal{L}_{0}$ the fields are evaluated at $y^{\sigma}$.

Let us write the Euler-Lagrange motion equations:\\

The equation for the particle,
\[\frac{\partial \mathcal{L}_{Tot}}{\partial x^{\sigma}}-\frac{d}{d\tau}\left(\frac{\partial \mathcal{L}_{Tot}}{\partial u^{\sigma}}\right)=0 \Longrightarrow \frac{\partial {\widehat{\mathcal{L}}}_{matt}}{\partial x^{\sigma}}-\frac{d}{d\tau}\left(\frac{\partial {\widehat{\mathcal{L}}}_{matt}}{\partial u^{\sigma}}\right)=0\,,\]

\noi results in

\be
g_{\mu\sigma}\frac{du^{\mu}}{d\tau}=u^{\mu}u^{\nu}(-\partial_{\nu}g_{\mu\sigma}+\frac{1}{2}\partial_{\sigma}g_{\mu\nu})-\frac{e}{m}u^{\mu}(\partial_{\sigma}A^{(ele)}_{\mu}-\partial_{\mu}A^{(ele)}_{\sigma})\
.
\ee

\noi Note that the Levi-Civita connection $\Gamma^{(L-C)\epsilon}_{\mu\nu}$ associated with the metric $g_{\mu\nu}$ is defined as usual, that is:

\[\Gamma^{(L-C)\epsilon}_{\mu\nu}=\frac{1}{2}g^{\epsilon\sigma}(\partial_{\mu}g_{\nu\sigma}+\partial_{\nu}g_{\mu\sigma}-\partial_{\sigma}g_{\mu\nu})\,,\]

\noi and

\[\Gamma^{(L-C)}_{\mu\nu,\epsilon}\equiv \Gamma^{(L-C)\epsilon}_{\mu\nu}g_{\epsilon\theta}=\frac{1}{2}(\partial_{\mu}g_{\nu\theta}+\partial_{\nu}g_{\mu\theta}-\partial_{\theta}g_{\mu\nu})\,.\]

On the other hand,

\[u^{\mu}u^{\nu}(-\partial_{\nu}g_{\mu\sigma}+\frac{1}{2}\partial_{\sigma}g_{\mu\nu})=-\frac{1}{2}u^{\mu}u^{\nu}(\partial_{\mu}g_{\nu\sigma}+\partial_{\nu}g_{\mu\sigma})+\frac{1}{2}u^{\mu}u^{\nu}\partial_{\sigma}g_{\mu\nu}=-u^{\mu}u^{\nu}\Gamma^{(L-C)}_{\mu\nu,\sigma}.\]

\noi Then, we arrive at:

\be
g_{\mu\sigma}\frac{du^{\mu}}{d\tau}=-u^{\mu}u^{\nu}\Gamma^{(L-C)}_{\mu\nu,\sigma}-\frac{e}{m}u^{\mu}(\partial_{\sigma}A^{(ele)}_{\mu}-\partial_{\mu}A^{(ele)}_{\sigma})\,.
\ee

\noi The last equation would be the usual motion equation for a particle in the presence of 
both gravitational and electromagnetic fields, although the difference manisfests itself 
in that the fields satisfy motion equations with mixing terms.

Decomposing the electromagnetic field to the first order in $\kappa$, 

\[A^{(ele)}_{\mu}=A^{(ele)0}_{\mu}+\kappa A^{(ele)1}_{\mu}\,,\]

\noi the motion equation for the particle reads:
\be
g_{\mu\sigma}\frac{du^{\mu}}{d\tau}=-u^{\mu}u^{\nu}\Gamma^{(L-C)}_{\mu\nu,\sigma}-\frac{e}{m}u^{\mu}(\partial_{\sigma}A^{(ele)0}_{\mu}-\partial_{\mu}A^{(ele)0}_{\sigma})-\frac{\kappa e}{m}u^{\mu}(\partial_{\sigma}A^{(ele)1}_{\mu}-\partial_{\mu}A^{(ele)1}_{\sigma})\,.\label{motion}
\ee

\noi This equation is compatible with that obtained in Ref.\cite{ egmixing} by group-theoretical 
methods, although, there,  the results were given up to a lower order in the group law, making the 
exact identification of terms by no means direct. Putting the approximate solution (\ref{A1}) into 
(\ref{motion}), the general comments after (\ref{A1}) do apply here.

As far as the field equations are concerned, and assuming that the Lagrangian density 
for the free compensating fields has the form 
\, $\Lambda \mathcal{L}_{0}\sim \Lambda (\mathcal{F}^{(ele)}_{\mu\nu}\mathcal{F}^{(ele)\mu\nu}+\mathcal{F}^{(\mu\nu)}_{\mu\nu})$,\,  the motion equations for the fields are similar to those of 
the previous section except for additional terms due to the presence of the particle:
\bea
&1)&C^{(ele)}_{\sigma,\epsilon\theta}A^{(\sigma)}_{\nu}F^{(ele)\mu\nu}+k^{\mu}_{\rho}T^{\rho}_{\epsilon\theta}-k^{\mu}_{\theta}T^{\rho}_{\epsilon\rho}+k^{\mu}_{\epsilon}T^{\rho}_{\theta\rho}+(k^{\mu}_{\rho}k^{\nu}_{\theta}-k^{\mu}_{\theta}k^{\nu}_{\rho}A^{(\rho}_{\epsilon)\nu}-(k^{\mu}_{\epsilon}k^{\nu}_{\rho}+k^{\mu}_{\rho}k^{\nu}_{\epsilon})A^{(\rho}_{\theta)\nu}=0
\ ,\nn\\
&2)&\partial_{\sigma}(\Lambda F^{(ele)\mu\sigma})=eu^{\mu} \ ,\nn\\
&3)&F^{(\nu\sigma)}_{\mu\sigma}-\frac{1}{2}\delta^{\nu}_{\mu}F^{(\sigma\lambda)}_{\sigma\lambda}=\mathcal{T}^{\nu}_{\mu(Tot)},\nn
\eea

\noi with

\[\mathcal{T}^{\nu}_{\mu(Tot)}\equiv \mathcal{T}^{\nu(mix)}_{\mu}+\mathcal{T}^{\nu(ele)}_{\mu}+\mathcal{T}^{\nu(matt)}_{\mu}\]

\noi defining the total energy-momentum tensor, where 

\[\mathcal{T}^{\nu(mix)}_{\mu}=-k^{\rho}_{\mu}\frac{\partial {\widehat{\mathcal{L}}}_{matt}}{\partial k^{\rho}_{\nu}}\]
\[\mathcal{T}^{\nu(ele)}_{\mu}=-F^{(ele)\nu}_{\sigma}F^{(ele)\sigma}_{\mu}+\frac{1}{2}\delta^{\nu}_{\mu}F^{(ele)}_{\sigma\lambda}F^{(ele)\sigma\lambda}\]
\[\mathcal{T}^{\nu(matt)}_{\mu}=\frac{1}{2}\delta^{\sigma}_{\mu}C^{ele}_{\sigma,\theta\epsilon}q^{\nu}_{\rho}F^{(ele)\rho\tau}A^{(\theta\epsilon)}_{\tau}.\]

\noi Similar comments to those made for equations (\ref{(1)}), (\ref{(2)}) and (\ref{(3)}) also apply
here.

Finally, we consider these equations to the first order in $\kappa$. In equation 1), the 
electromagnetic tensor is already accompanied by the constant $C^{(ele)}_{\sigma,\epsilon\theta}$, 
which is proportional to $\kappa$, so that we must replace $F^{(ele)\mu\nu}$ with the unmixed 
electromagnetic tensor in order to obtain an equation for the Lorentz potential to first order 
in~$\kappa$.

With respect to equation 2) we have:\\

{\it Zero order}:

\bea
\partial_{\sigma}[\Lambda g^{\mu\epsilon}g^{\sigma\theta}(A^{(ele)0}_{\epsilon,\theta}-A^{(ele)0}_{\theta,\epsilon})]=eu^{\mu}
 \ .\eea

\noi $A^{(ele)0}_{\mu}$ is the usual electromagnetic field without mixing.\\

{\it First order}:

\bea
\partial_{\sigma}\{\Lambda g^{\mu\epsilon}g^{\sigma\theta}[A^{(ele)1}_{\epsilon,\theta}-A^{(ele)1}_{\theta,\epsilon}-\frac{1}{2}M^{ele}_{\rho,\nu\xi}(A^{(\rho)}_{\epsilon}A^{(\nu\xi)}_{\theta}-A^{(\rho)}_{\theta}A^{(\nu\xi)}_{\epsilon})]\}=0.
\eea

As in equation (\ref{(333)}), up to first order in $\kappa$, the Lorentz potentials coincide with the Lorentz potentials of the gauge theory of Poincar\'e group in vacuum.

\section{Outlook}

The possibility of formulating the Lagrangian version of the already proposed revision of the Higgs-Kibble mechanism, in a group approach to the quantization of Yang-Mills theories \cite{chorizo}, to provide mass to the massive vector bosons, using only the group parameters without extra fields, is also expected to be carried out in the near future.


\noi Related to the unification between gravity and the internal gauge interactions
it is also known that a semidirect action of the group of diffeomorphisms on the gauge group 
\[\mathcal{D}iff(M)\otimes^{s}G(M)\] exists and    that given a gauge group $G(M)$ it is possible to find a formal group composition law for $G(M)$. Therefore we can also find  a group composition law for that semidirect product of groups, what implies a non-trivial mixture between gravity and the internal interactions. This new mixing just proposed here is quite different from the one described in this work and deserves further study.

\section*{Appendix. Lagrangian Formalism on $J^{1}(J^{1}(E))$}

The space $J^{1}(J^{1}(E))$ is the bundle of the 1-jets associated with the fibre space $(J^{1}(E),M,\pi^{1})$. The coordinate system for 
$J^{1}(J^{1}(E))$ will be defined by

\bea
(x^{\mu},y^{\alpha},y^{\alpha}_{\mu};z^{\alpha}_{,\nu},z^{\alpha}_{\mu,\nu})
\ .
\eea
A Lagrangian density of type $\mathcal{L}^{1,1}$ is an application of  $J^{1}(J^{1}(E))$ on R.

The 1-jet prolongation of $\varphi^{1}$, $j^{1}(\varphi^{1})\equiv \bar{\varphi^{1}}$ is the only section of
$(J^{1}(J^{1}(E)),M,\pi)$ such that $j^{1}$ is an injection of $\Gamma(J^{1}(E))$ into $\Gamma(J^{1}(J^{1}(E)))$ and
\bea
\theta^{1,1\alpha}|_{\bar{\varphi^{1}}}=0\\
\theta^{1,1\alpha}_{\mu}|_{\bar{\varphi^{1}}}=0 \ ,
\eea
where the structure forms $\theta^{1,1}$ are now

\bea
\theta^{1,1\alpha}=dy^{\alpha}-z^{\alpha}_{,\nu}dx^{\nu}\\
\theta^{1,1\alpha}_{\mu}=dy^{\alpha}_{\mu}-z^{\alpha}_{\mu,\nu}dx^{\nu} \
.
\eea

The jet extension (prolongation) condition requires
\bea
z^{\alpha}_{,\nu}=\partial_{\nu}y^{\alpha}\\
z^{\alpha}_{\mu,\nu}=\partial_{\nu}y^{\alpha}_{\mu} \ .
\eea 

Given a vector field on $J^{1}(E)$, $X^{1} \in \Gamma(T(J^{1}(E)))$, its 1-jet extension (or prolongation) through the injection $j^{1}$ is the only field $j^{1}(X^{1})\equiv \bar{X^{1}}$ such that it is an infinitesimal contact transformation, i.e.,

\bea
L_{\bar{X^{1}}}\theta^{1,1\alpha}=A^{\alpha;}_{;\beta}\theta^{1,1\beta}+A^{\alpha;\nu}_{;\beta}\theta^{1,1\beta}_{\nu}\\
L_{\bar{X^{1}}}\theta^{1,1\alpha}_{\mu}=A^{\alpha;}_{\mu;\beta}\theta^{1,1\beta}+A^{\alpha;\nu}_{\mu;\beta}\theta^{1,1\beta}_{\nu}
\ ,
\eea
where the 1-jet extension of $X^{1}$ is determined from the general form of $X^{1,1}$,

\bea
X^{1,1}=X^{\mu}\frac{\partial }{\partial x^{\mu}}+X^{\alpha}\frac{\partial }{\partial y^{\alpha}}+
X^{\alpha}_{\mu}\frac{\partial }{\partial y^{\alpha}_{\mu}}+X^{\alpha}_{,\mu}\frac{\partial }{\partial z^{\alpha}_{,\mu}}+X^{\alpha}_{\mu,\nu}\frac{\partial }{\partial z^{\alpha}_{\mu,\nu}} \ .
\eea 

Hence

\bea
A^{\alpha;}_{;\beta}=\frac{\partial X^{\alpha}}{\partial y^{\beta}}-z^{\alpha}_{,\sigma}\frac{\partial X^{\sigma}}{\partial y^{\beta}}\\
A^{\alpha;}_{\mu;\beta}=\frac{\partial X^{\alpha}_{\mu}}{\partial y^{\beta}}-z^{\alpha}_{\mu,\sigma}\frac{\partial X^{\sigma}}{\partial y^{\beta}}\\
A^{\alpha;\nu}_{;\beta}=\frac{\partial X^{\alpha}}{\partial y^{\beta}_{\nu}}-z^{\alpha}_{,\sigma}\frac{\partial X^{\sigma}}{\partial y^{\beta}_{\nu}}\\
A^{\alpha;\nu}_{\mu;\beta}=\frac{\partial X^{\alpha}_{\mu}}{\partial y^{\beta}_{\nu}}-z^{\alpha}_{\mu,\sigma}\frac{\partial X^{\sigma}}{\partial y^{\beta}_{\nu}}
\eea
and the components $\bar{X}^{\alpha}_{,\mu}$, $\bar{X}^{\alpha}_{\nu,\mu}$ read:

\bea
\bar{X}^{\alpha}_{,\mu}=\frac{\partial X^{\alpha}}{\partial x^{\mu}}-z^{\alpha}_{,\sigma}\frac{\partial X^{\sigma}}{\partial x^{\mu}}+
A^{\alpha;}_{;\beta}z^{\beta}_{,\mu}+A^{\alpha;\sigma}_{;\beta}z^{\beta}_{\sigma,\mu}\\
\bar{X}^{\alpha}_{\nu,\mu}=\frac{\partial X^{\alpha}_{\nu}}{\partial x^{\mu}}-z^{\alpha}_{\nu,\sigma}\frac{\partial X^{\sigma}}{\partial x^{\mu}}+
A^{\alpha;}_{\nu;\beta}z^{\beta}_{,\mu}+A^{\alpha;\sigma}_{\nu;\beta}z^{\beta}_{\sigma,\mu}
\ .
\eea

Given a Lagrangian density $\mathcal{L}^{1,1}$, the Hamilton functional $\mathcal{S}^{1,1}$ is the application of $\Gamma(J^{1}(E))$ on R defined by

\bea
\mathcal{S}^{1,1}(\varphi^{1})=\int_{j^{1}(\varphi)(M)}\mathcal{L}^{1,1}(j^{1}(\varphi^{1}))\omega
\eea

\noi $\forall \varphi^{1} \in \Gamma(J^{1}(E)).$

The Hamilton Principle states that the trajectories of the variational problem are the solutions of

\bea
\int_{j^{1}(\varphi^{1})(M)}L_{\bar{X^{1}}}(\mathcal{L}^{1,1}(j^{1}(\varphi^{1}))\omega))=0
\eea

\noi $\forall X^{1} \in \Gamma(T(J^{1}(E))).$

This principle leads to a set of Euler-Lagrange equations of the form

\bea
\frac{\partial \mathcal{L}^{1,1}}{\partial y^{\alpha}}-\frac{d}{dx^{\mu}}\left(\frac{\partial \mathcal{L}^{1,1}}{\partial z^{\alpha}_{,\mu}}\right)=0\\
\frac{\partial \mathcal{L}^{1,1}}{\partial y^{\alpha}_{\nu}}-\frac{d}{dx^{\mu}}\left(\frac{\partial \mathcal{L}^{1,1}}{\partial z^{\alpha}_{\nu,\mu}}\right)=0 \ .
\eea

Given a $\mathcal{L}^{1,1}$, the Poincar\'e -Cartan form is given by

\bea
\Theta^{1,1}_{PC}=\mathcal{L}^{1,1}\omega+\theta^{1,1\alpha} \wedge \Omega_{\alpha}+\theta^{1,1\alpha}_{\mu} \wedge \Omega_{\alpha}^{\mu} \ ,
\eea

where 

\bea
\Omega_{\alpha}\equiv \frac{\partial \mathcal{L}^{1,1}}{\partial z^{\alpha}_{,\nu}}\theta_{\nu}\\
\Omega_{\alpha}^{\mu}\equiv \frac{\partial \mathcal{L}^{1,1}}{\partial z^{\alpha}_{\mu,\nu}}\theta_{\nu}\\
\theta_{\mu}\equiv (-)^{\mu}dx^{0} \wedge...\wedge d\hat{x}^{\mu} \wedge...\wedge dx^{3} \ .
\eea

Given $\Theta^{1,1}_{PC}$ on $J^{1}(J^{1}(E))$ the modified Hamilton functional $\mathcal{S}'^{1,1}$ is the application of $\Gamma(J^{1}(J^{1}(E)))$ on R defined by 

\bea
\mathcal{S}'^{1,1}(\varphi^{1,1})\equiv \int_{\varphi^{1,1}(M)}\Theta^{1,1}_{PC}
\eea

\noi $\forall \varphi^{1,1} \in \Gamma(J^{1}(J^{1}(E)))$

The Modified Hamilton Principle states that a cross section is critical if and only if

\bea
(\delta \mathcal{S}'^{1,1})_{\varphi^{1,1}}(X^{1,1})=\int_{\varphi^{1,1}(M)}L_{X^{1,1}}\Theta^{1,1}_{PC}=0
\eea

\noi $\forall X^{1,1} \in \Gamma(T(J^{1}(J^{1}(E)))$

The critical sections are thus the solutions of

\bea
i_{X^{1,1}}d\Theta^{1,1}_{PC}|_{\varphi^{1,1}}=0
\eea
from where the system of equations which come from the coefficients of $X^{\beta}_{\sigma,\nu}$ and $X^{\beta}_{,\nu}$ read respectively:

\bea
\frac{\partial }{\partial z^{\beta}_{\sigma,\nu}}\frac{\partial \mathcal{L}^{1,1}}{\partial z^{\alpha}_{,\mu}}\left(\frac{\partial y^{\alpha}}{\partial x^{\mu}}-z^{\alpha}_{\mu}\right)+\frac{\partial }{\partial z^{\beta}_{\sigma,\nu}}\frac{\partial \mathcal{L}^{1,1}}{\partial z^{\alpha}_{\rho,\mu}}\left(\frac{\partial y^{\alpha}_{\rho}}{\partial x^{\mu}}-z^{\alpha}_{\rho,\mu}\right)=0\\
\frac{\partial }{\partial z^{\beta}_{,\nu}}\frac{\partial \mathcal{L}^{1,1}}{\partial z^{\alpha}_{,\mu}}\left(\frac{\partial y^{\alpha}}{\partial x^{\mu}}-z^{\alpha}_{\mu}\right)+\frac{\partial }{\partial z^{\beta}_{,\nu}}\frac{\partial \mathcal{L}^{1,1}}{\partial z^{\alpha}_{\rho,\mu}}\left(\frac{\partial y^{\alpha}_{\rho}}{\partial x^{\mu}}-z^{\alpha}_{\rho,\mu}\right)=0 \ .
\eea

\noi When this system admits the trivial solution (regularity condition for $\mathcal{L}^{1,1}$), the 1-jet prolongation condition is satisfied, and then the remaining equations (from the coefficients of $X^{\alpha}_{\nu}$ and $X^{\alpha}$ respectively)

\bea
-\frac{d}{dx^{\mu}}\left(\frac{\partial \mathcal{L}^{1,1}}{\partial z^{\alpha}_{\nu,\mu}}\right)+
\frac{\partial \mathcal{L}^{1,1}}{\partial y^{\alpha}_{\nu}}+
\frac{\partial }{\partial y^{\alpha}_{\nu}}\left(\frac{\partial \mathcal{L}^{1,1}}{\partial z^{\beta}_{,\mu}}\right)
\left(\frac{\partial y^{\beta}}{\partial x^{\mu}}-z^{\beta}_{,\mu}\right)+
\frac{\partial }{\partial y^{\alpha}_{\nu}}\left(\frac{\partial \mathcal{L}^{1,1}}{\partial z^{\beta}_{\sigma,\mu}}\right)
\left(\frac{\partial y^{\beta}_{\sigma}}{\partial x^{\mu}}-z^{\beta}_{\sigma,\mu}\right)=0\\
-\frac{d}{dx^{\mu}}\left(\frac{\partial \mathcal{L}^{1,1}}{\partial z^{\alpha}_{,\mu}}\right)+
\frac{\partial \mathcal{L}^{1,1}}{\partial y^{\alpha}}+
\frac{\partial }{\partial y^{\alpha}}\left(\frac{\partial \mathcal{L}^{1,1}}{\partial z^{\beta}_{,\mu}}\right)
\left(\frac{\partial y^{\beta}}{\partial x^{\mu}}-z^{\beta}_{,\mu}\right)+
\frac{\partial }{\partial y^{\alpha}}\left(\frac{\partial \mathcal{L}^{1,1}}{\partial z^{\beta}_{\sigma,\mu}}\right)
\left(\frac{\partial y^{\beta}_{\sigma}}{\partial x^{\mu}}-z^{\beta}_{\sigma,\mu}\right)=0
\eea

\noi reproduce the Euler-Lagrange equations of the Ordinary Hamilton Principle and thus both principles are equivalent.


\end{document}